\documentclass[authoryear,final,12pt,times]{elsarticle}
 \usepackage[margin=2.50cm]{geometry}
 \usepackage{graphics}
\usepackage{epsfig}
\usepackage[breaklinks=true]{hyperref}
\usepackage{breakcites}
 \usepackage{subfigure}
 \usepackage{epsfig}
\usepackage[cmex10]{amsmath}
 \usepackage{booktabs}
\usepackage{verbatim} 
 \usepackage{epstopdf}
\usepackage{tikz}
\usepackage{mathtools}
\usepackage{array} 
\usepackage{amssymb}
 \usepackage{amsthm}
\usepackage{lineno}
    \makeatletter
    \def\ps@pprintTitle{%
      \let\@oddhead\@empty
      \let\@evenhead\@empty
      \let\@oddfoot\@empty
      \let\@evenfoot\@oddfoot
    }
    \makeatother
\journal{Journal Name}
\begin{document}
\renewcommand{\baselinestretch}{1.0}
 \begin{sloppypar}
\begin{frontmatter}
\title { Modified Group Delay Based MultiPitch Estimation in Co-Channel Speech} 
%% use optional labels to link authors explicitly to addresses:

\author{  Rajeev Rajan, Hema A. Murthy}
 \address{Department of Computer Science and Engineering \\
 Indian Institute of Technology Madras \\
Chennai,  India\\
E-mail:\{ rajeev,hema \} @cse.iitm.ac.in } 
 \begin{abstract}
%% Text of abstract
% In this paper, we propose a multi-pitch estimation algorithm for concurrent speech based on modified group delay functions.  As the phase is difficult to process, 
% the group delay function,  which can be computed directly from the speech signal,  has been
% widely used  to extract source and system parameters. 
Phase processing has been replaced by group delay processing for the extraction of source and system parameters from speech.  
Group delay functions are ill-behaved when the transfer function has zeros that are close to unit circle in the $z$-domain.    The modified group delay function addresses this problem and has been successfully used for formant and monopitch estimation.   
In this paper, modified group delay functions are used for  multipitch estimation in concurrent speech. 
The power spectrum of the speech is first flattened in order to annihilate
the system characteristics,  while retaining  the source characteristics.  Group delay analysis on this flattened spectrum picks
the predominant pitch in the first pass 
and a comb filter is used to filter out the estimated pitch along with its harmonics.  The residual spectrum  is again analyzed for the next
candidate pitch estimate in the second pass. The final pitch trajectories of the constituent speech utterances are formed using pitch grouping and post processing techniques.  The performance of the proposed algorithm was evaluated on standard datasets  using two metrics; pitch accuracy and standard deviation of fine pitch error.
Our results show that the proposed algorithm is a promising  pitch detection method in multipitch environment for real speech recordings.  
\end{abstract}
\end{frontmatter}
\section{Introduction}
\label{Background}
In speech and music research, robust pitch detection is a fundamental problem which finds many
applications in day to day life. Pitch is the auditory attribute of a sound that allows its ordering on a frequency related scale. The rising and falling of pitch contours help in conveying prosody in speech and in tone languages, 
determine the meaning of words \citep{neuro}.  A detail review
on various monopitch estimation algorithms can be seen in \citep{hess,rabinerPitch,gerhard}. Pitch detection algorithms can be broadly classified into methods which 
operate in time domain, frequency domain, or both. 
The most commonly used time domain approaches are autocorrelation function and average magnitude difference function. In the frequency domain approaches, locating harmonic 
peaks is the key step in  most of the algorithms \citep{Schroeder}. Studies show  that in tonal languages, the relative pitch motion of an utterance
contributes to the lexical information contained in a word unit \citep{gerhard}.  We cannot ignore the pitch information during recognition in such instances. 
 A majority of the pitch tracking methods are usually limited to  clean speech and give 
a degraded performance in the presence of other speakers or  noise.  When a combination of speech utterances from two or more speakers are  transmitted through a  single channel, pitch cues of the individual sources 
will be weakened by the presence of mutual interference.  In such ambiguous situations, estimating the accurate  pitch tracks  is a challenging task and currently is far from 
being completely solved, despite the attempts of several state-of-the-art approaches.

The multi-pitch estimation problem can be formulated as follows \citep{chrsitensen2}: Consider
a signal consisting of several, say ${\small{K}}$, sets of harmonics
 with fundamental frequencies  $\omega_k $,
 for $k$ = 1, \dots,$K$, that is corrupted by an additive white
Gaussian noise $\omega[n]$, having
variance $\sigma^2$, for n = 0, . . . , $N-1$, i.e.,
\begin{equation}
\label{harmonic}
 x[n] = \sum_{k = 1}^{K} { \sum_{l = 1}^{ L} {a_{k,l} e^{j\omega_k ln} + {\omega[n]}}}
\end{equation}

where $ a_{k,l} = A_{k,l}e^{j\phi_{k,l}} $ is the complex amplitude of the $ l^{th}$ harmonic of the source with $A_{k,l} > 0 $, 
 $\phi_{k,l}$  being the amplitude
and the phase of the $ l^{th}$  harmonic of the $ k^{th}$ source 
respectively. The model in Equation (\ref{harmonic}) is known as the harmonic sinusoidal model. The
task is to estimate the individual pitch estimates ${\omega_k}$ in the mixture signal. 
                 The estimation of the fundamental frequency, or the pitch of 
audio signals has a wide range of applications in Computational Auditory Scene Analysis (CASA), prosody analysis, source separation and speaker identification \citep{cheveigne,hamsadhana}. 
In music also, multipitch estimation is inevitable in applications such as the extraction of ``predominant $F_o$''\citep{justin}, computation of  bass line \citep{Goto}, content-based indexing of audio databases \citep{content} and automatic transcription \citep{matti}.
Note that the interactive music applications demand highly robust real time pitch estimation algorithms in  all aspects.   
\section{Related work}
Numerous methods have been reported for multipitch estimation in speech and music \citep{li1,kameoka,Wu}. % In speech multi-pitch tracking algorithms \citep{ref}  are proposed mainly for  accurate pitch detection in noisy and reverberant 
% environments and co channel speech separation. 
 The correlogram based algorithm proposed by Wu {\it {et al.}} \citep{Wu} uses a unitary model 
of pitch perception to estimate the pitch of multiple speakers.  The  input signal 
is decomposed into sub-bands using a gammatone filterbank and the framewise normalized autocorrelation function is computed for each channel.  The peaks selected from all the channels are used to compute a likelihood  of pitch periodicities and these likelihoods are modeled by a Hidden Markov
Model (HMM) to generate the  pitch trajectories.  A subharmonic summation method and a spectral cancellation framework is used in the co-channel speech separation algorithm proposed by Li {\it{et al.}} \citep{li1}.
Multi-pitch trajectory estimation based on  harmonic Gaussian Mixture Model (GMM) and  nonlinear Kalman filtering is also proposed for multipitch environments \citep{kameoka2}. 
A constrained GMM based approach on the platform of information criterion is attempted in  \citep{kameoka}.

In a polyphonic context, the overlap between the overtones of different notes and the unknown number of notes occurring simultaneously make the multipitch 
estimation a difficult and challenging task \citep{emiya}.  The algorithms used in polyphonic environment for  pitch transcription  include  auditory scene analysis based methods \citep {tanaka,mellinger}, 
signal model based Bayesian inference methods \citep{goto2}, unsupervised learning methods \citep{smargdis,virtanen} 
and auditory model based methods \citep{klappuri3,tolonen,Wu}.   In auditory scene analysis based methods, acoustic features and musical information are used to group the sound sources present in a scene, while signal model based methods employ parametric signal models and statistical methods to transcribe the pitch tracks. Unsupervised learning techniques include independent
component analysis, non-negative matrix factorization, usage of source-specific prior knowledge  and sparce coding.   In auditory model based methods, a peripheral hearing model is 
used  for the intermediate data representation of the mixture signal, followed by periodicity analysis and iterative cancellation.
Multi-pitch estimation in music can be used to extract 
various information such as number of 
simultaneous sounds, spectral envelopes and onset time/offset time of notes.  If the pitch of a sound 
can be determined  without getting confused by other co-occurring sounds, the pitch information can be used to organize simultaneous spectral components
for their production \citep{kla2}.

Although the pitch is based on timing, it is hardly exploited for pitch estimation, 
primarily because phase appears to be noisy owing to the wrapping problem.   On the other hand, group delay 
function that preserves the properties of the phase can be exploited.  
Group delay functions are poorly behaved when the signal is nonminimum phase.   
The modified group delay function was proposed in \citep{Gd1} to address this issue.  
In this paper, this idea is extended to multipitch analysis.
We propose a phase based signal processing algorithm  as opposed to  
conventional magnitude based methods to retrieve the individual 
pitches in concurrent speech.  The phase spectrum has to be first
unwrapped before any meaningful analysis can be
 performed.  The advantage of the group delay function instead of the phase spectrum is that it can be computed directly from the signal.  Hence 
 the problem of unwrapping of the phase spectrum can be solved.
 
 The primary motivation for this work arises
from the applications of the group delay function in
estimating sinusoids from noise \citep{Gd1}. The algorithm starts from the flattened power spectrum.
% In speech production,
% the source can be approximately modeled by a periodic
% function (that is approximately a sinusoid) that modulates
% the formants.
% If the carrier frequencies (or rather formants) are suppressed
% in the power spectrum, the periodic source will result in a
% periodic function in the spectrum. 
The modified power
spectrum can be thought as a sum of sinusoids.  This is then subjected to modified group delay processing 
to estimate the pitch components present in the speech mixture by iterative estimation and cancellation.
% modified power spectrum can then be subjected to sinusoidal
% analysis.  
Group delay based pitch extraction for a single voice is described in \citep{hamicassp1991}.

 \indent The outline of the rest of paper is as follows. Section \ref{sec:gd} explains group delay functions and modified group delay function briefly.  The
theory of pitch detection using modified group delay functions is described in Section  \ref{theory}. In Section \ref{multi1}, the proposed system for  multi-pitch estimation is discussed in detail.   Section  \ref{dataset} discusses the dataset and evaluation metrics followed by  results and analysis in Section \ref{results}. 
The effectiveness of a variant of group delay feature is explained in Section \ref{smcc}. Conclusions are finally drawn in Section \ref{sec:conclusion}.

%\section{Preliminaries}
\section{Group-delay functions and modified group delay functions ({\small{MODGD}})}
\label{sec:gd}
\indent Signals  can be represented in different domains such as time domain, frequency domain, z-domain
 and cepstral domain.  In \citep{hemathesis}, it was shown that signal information can  be represented by 
group delay functions,  one derived from the  magnitude of the Fourier transform and the other from the Fourier transform phase. 
%Group delay function $\tau(\omega)$ is defined as 

Consider a discrete time signal  $  x[n] $. Then
\begin{equation}
\label{eq:eqn2}
 X(e^{j\omega})=  |X(e^{j\omega})| e^{j \arg(X(e^{j\omega})) }
\end{equation}
 where $X(e^{j\omega})$ is the Fourier Transform (FT) of the signal ${x[n]}$
and $ \arg(X(e^{j\omega})) $ is the phase function.

The group delay function $\tau(e^{j\omega})$ is defined as the negative derivative
of  the unwrapped Fourier transform phase with respect to the frequency.
\begin{equation}
\label{eq:eqn1}
\tau(e^{j\omega}) =-\frac{d \{\arg(X(e^{j\omega})) \} }{d\omega}
\end{equation}
From Equation (\ref{eq:eqn2})
\begin{equation}
\label{eq:eqn4}
\arg(X(e^{j\omega})) = Im[\log~ X(e^{j\omega})]
\end{equation}
Using  Equation (\ref{eq:eqn1}) and  Equation (\ref{eq:eqn4}), the group delay
function can be  computed directly  from the signal as shown below
\citep{oppenheim}:
\begin{equation}
\label{eq:eqn5}
\tau(e^{j\omega}) = -Im\frac{d(\log(X(e^{j\omega})))}{d\omega}
\end{equation}
\begin{equation}
\label{gdequation}
 \tau(e^{j\omega}) = \frac{X_R(e^{j\omega})Y_R(e^{j\omega})
+Y_I(e^{j\omega})X_I(e^{j\omega})}{| X(e^{j\omega})|}
\end{equation}
where the subscripts \emph{R} and \emph{I} denote the real and imaginary parts. $X(e^{j\omega})$  and $Y(e^{j\omega})$
are the Fourier transforms of x[n]and nx[n] respectively.

% \subsection{Modified Group-delay functions({\small{MODGD}})}
\indent It is important to note that the denominator term
$|X(e^{j\omega})| ^2 $ in Equation (\ref{gdequation}) becomes very small at zeros that are located
close to the unit circle.   This makes the group delay function
very spiky in nature and 
also alters the dynamic range of the
group delay spectrum.  As the spikiness of the group delay function has no role to play in source/system characteristics,
the computation of the group delay function is modified such that the source and system characteristics are not lost. The spiky nature of the group delay
spectrum can be overcome by replacing the term $ |X(e^{j\omega})|$ in
the denominator of the group delay function  with 
its cepstrally smoothed version, $S(e^{j\omega})$.
The new function obtained is referred to as the modified group delay function in the literature.
% and forms modified group delay function in Equation (\ref{mgdequation}).
The  algorithm for computation of the modified group delay function is described in \citep{hegde2007b} and is given as
\begin{equation}
\label{mgdequation}
\tau_m(e^{j\omega})  = \frac{X_R(e^{j\omega})Y_R(e^{j\omega})
+Y_I(e^{j\omega})X_I(e^{j\omega})}{ |S(e^{j\omega})|^{2\gamma}}
\end{equation}
where $S(e^{j\omega}) $ is the cepstrally smoothed version of
$X(e^{j\omega})$. The algorithm for the computation of MODGDF  is
given in \citep{hamsadhana}.  Two new parameters, $\alpha$
and $\gamma$  are introduced to control the dynamic range of 
MODGDF such that  0 $<$ $\alpha$  $\leq$  1 and  0 $<$ $\gamma$ $\leq$ 1.
Modified group delay based algorithms can be used  effectively  
to estimate system  and source characteristics in speech processing \citep{hamsadhana}. 

% If a signal is periodic,  it suggests that some basic waveform repeats with a certain frequency. 
% Mathematically, we can express it as $x[n]\approx x[n-D]$ 
% where $D$ is the repetition or pitch period.  A measure of periodicity can be obtained  using a metric  $e(n)$ defined as 
% 
% 
% where the constant $\alpha \in C$ is introduced to allow for some variation of the amplitude.
% 
% 
% In $Z$ domain,
% 
% 
% which shows that the process of matching of a signal by a delayed version of itself is  a filtering problem.  

\section{Theory of pitch detection using modified group delay functions}
\label{theory}
The vocal tract system and its excitation contribute to the envelope and the fine structure respectively of the speech spectrum.
The periodicity of the source manifests as picket fence harmonics in the power spectrum of the signal.  If the vocal tract  information can be suppressed, the picket fence harmonics
are essentially pure sinusoids.   % In speech production, the source can be approximately
%As shown in the block diagram, the power spectrum of the mixed speech is  first flattened.
\begin{figure}[h!]
\centering
\includegraphics[width=10cm,height=10cm]{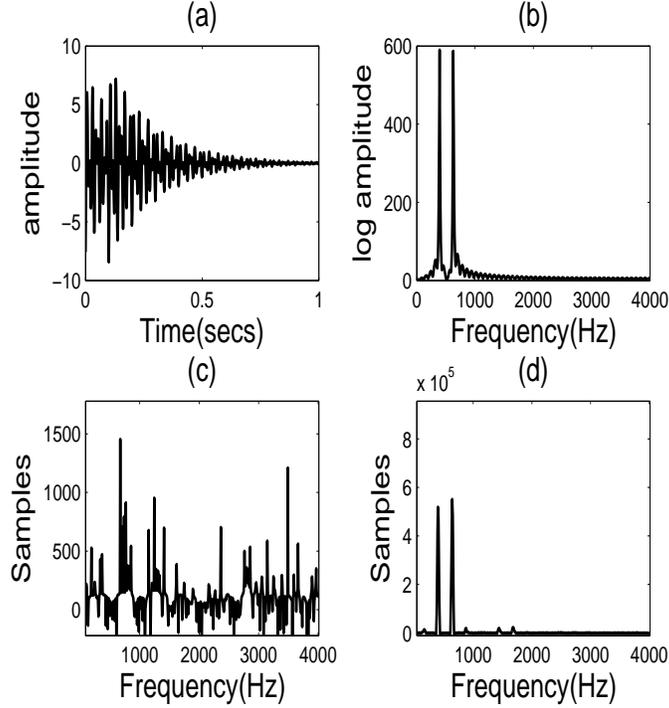}
\caption{\small{(a) Composite noisy signal,  (b) Magnitude spectrum of  a frame,   (c) Group delay corresponds to frame in (b),   (d) Modified group delay corresponds to frame in (b) }}
\label{fig_noisy}
\end{figure}
The modified power spectrum can be thought as a sinusoidal
signal. In the literature, it was shown that the modified group delay 
function is quite effective in estimating sinusoids in noise \citep{Gd1}.
High resolution property of modified group delay \citep{hemathesis,rajeshthesis} is exploited in all those cases to resolve the spectral components.
For instance, consider a noisy composite signal shown in  Figure \ref{fig_noisy}(a).   Figure \ref{fig_noisy}(b)  shows its magnitude spectrum. 
Even though the group delay is spiky in nature
(ref:-Figure \ref{fig_noisy}(c)),  spectral components are well resolved in  the {\small{MODGD}} feature space in Figure \ref{fig_noisy}(d). 
The monopitch estimation based on group delay function is explained in \citep{hemathesis}. The process is illustrated in Figure \ref{figxy} using the plots obtained in the intermediate steps.
A frame of speech is shown in Figure \ref{figxy}(a).  The flattened spectrum of the corresponding frame 
is shown in Figure \ref{figxy}(b).  Peaks at multiples of  fundamental frequencies can be observed in the {\small{MODGD}}
plot shown in Figure \ref{figxy}(c). The peak in the MODGD feature space in the range corresponds to $[P_{min}, P_{max}]$ is mapped to the 
pitch estimate.  The estimated pitch trajectory along with reference for an entire speech utterance is given in 
Figure \ref{figxy}(d). The systematic evaluation shows that the group delay based approach is at par with any other  magnitude based 
approaches \citep{hemathesis}.

\begin{figure}[h!]
\label{figxy}
\centering
\includegraphics[width=10cm,height=10cm]{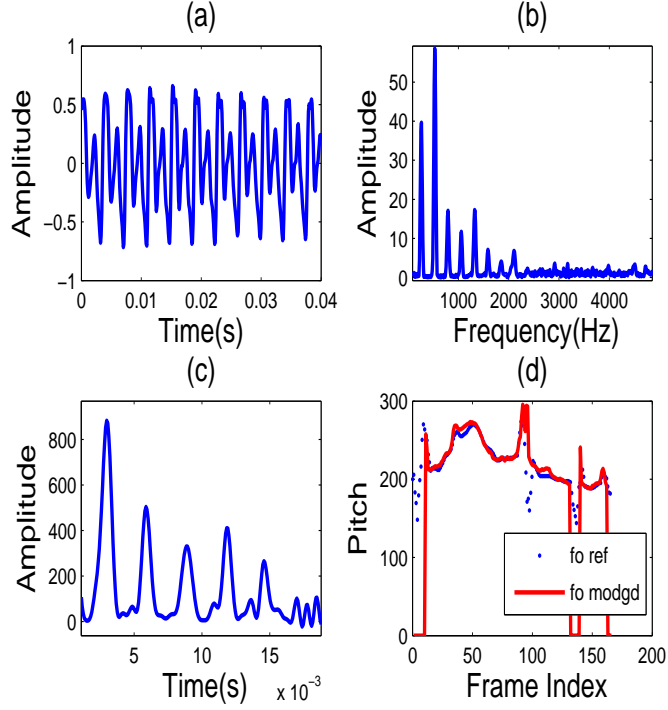}
\caption{\small{(a) Frame of a speech (b) Flattened power spectrum  (c) Peaks in the {\small{MODGD}} feature space (d) Pitch estimated for the entire utterance with reference}}
\label{figxy}
\end{figure}
%\subsection{Postprocessing}

The proposed method is an extension of the aforesaid process to multipitch environment. In the case of multiple speakers, the flattened power spectrum contains the excitation information of all the speakers.  
% This modified power spectrum can  be subjected
% to sinusoidal analysis  to estimate the pitch component present in the mixture.
%  
For instance, consider the  $z$-transform of  impulses separated by $T_o $ and $T_1 $, corresponds to the excitation components, then
\begin{equation}
    E(z)  = 1+ z^{-T_o}+ z^{-T_1} + z^{-2T_o} +  z^{-2T_1}
\end{equation}

The power spectrum of the source is given by
\begin{multline}
    E(z) E^*(z)  = (1+ z^{-T_o}+ z^{-T_1} + z^{-2T_o} +  z^{-2T_1})
    (1 + z^{ T_o} + z^{T_1} + z^{2T_o} +  z^{2T_1})
\end{multline}
Substituting $z = e^{j\omega} $,
{\small{\begin{multline}
    \mid E(e^{j\omega})\mid ^2= 5 + 4 \cos(\omega{T_o})+ 4 \cos(\omega{T_1})+ \\
    2 \cos(\omega{2T_o})+ 
 2 \cos(\omega{2T_1})+ 2\cos(\omega({T_o-2T_1}))+ \\ 2\cos(\omega({T_1-2T_0})) + 2\cos(\omega(2({T_1-T_0}))) + 2\cos(\omega({T_1-T_0}))
  \end{multline}}}

By restricting to three impulses per frame and evaluating the power spectrum on the
unit circle as above and introducing a parameter $\gamma$, we have
{\small{\begin{equation}
 \label{eq:flatspec}
 \mid E(e^{j\omega})\mid ^{2^\gamma} = (3 + 2(1+ \cos (\omega T_o)+ \cos (\omega T_1)+ \cos (\omega(T_0- T_1) )^\gamma
\end{equation}}}

where $0 < \gamma \leq 1$.   The parameter $\gamma$ controls the
flatness of the spectrum.  Thus the signal is a sum of sinusoids with frequencies that are
integral multiples of $\frac{1}{T_o}$ ,$\frac{1}{T_1}$, $\frac{1}{T_0-T_1}$ and few combinations.   If the spectral components corresponding to the periodic component are emphasised, the problem
of pitch extraction reduces to that of the estimation of sinusoids in the frequency domain.  We now replace $\omega$ by n  and  $T_o$, $T_1$ by $\omega_o$, $\omega_1$ in Equation (\ref{eq:flatspec}) and remove the dc component  to obtain a signal which is ideally
 a sum of sinusoids corresponds to excitation components present in the mixture.
\begin{multline}
\label{spectrum1}
    s[n] =\cos(n\omega_o) + \cos(n\omega_1) +  \cos n(\omega_1 - \omega_0) \\
    +cos(n 2 \omega_o)+ \cos(n 2\omega_1) .. \\
    n = 0,1,2,3.......N-1
\end{multline}
This signal is subjected to modified group delay processing,  which results in peaks at multiples of the  partials present in the speech mixture. The procedure to map this peak locations to
constituent pitch trajectories is explained in the next section.
% We first
% flatten the power spectrum usin ., i.e.g a root cepstral based smoothing technique\citep{hamicassp1991}.
% The high signal to noise ratio regions of this spectrum of Equation (\ref{spectrum}) are processed using
% the modified group delay function to detect the spectral components. 

% estimate with abrupt change with previous values  If detected, it is
% replaced by interpolated sequence, maintaining pitch continuity and followed by median filtering. 

\section{Proposed system description}
\label{multi1}
\begin{figure}[h!]
\centering
\includegraphics[width=12cm,height=8cm]{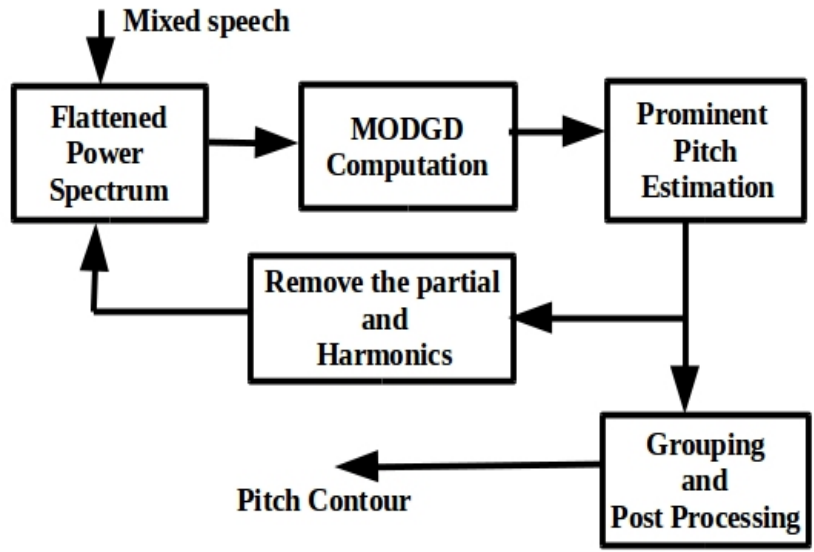}
\caption{\small{Block diagram of the proposed method}}
\label{blockdiagram}
\end{figure}

The block diagram of the  proposed system is shown in Figure \ref{blockdiagram}.  As seen in the Figure, the power spectrum of the speech signal is first flattened 
using cepstral smoothing technique to annihilate system characteristics by retaining the  excitation information.
In the mixed speech, the flattened spectrum consists of excitations of both the speakers. The flattened spectrum is 
frame-wise analysed using {\small {MODGD}}  algorithm described in Section \ref{sec:gd}.  As discussed in the Section \ref{theory}, peaks can be seen  in the  { \small{MODGD}} feature space at locations corresponding to the multiples of
all the pitch components and its few algebraic combinations.
%Strong repetitive  pitch track pattern can be observed
% at the multiples of fundamental frequencies in the modgdgram\footnote{ A modgdgram is a visual representation 
% of the modified group delay function of a speech utterance as it varies with time.
% Modified group delay function at particular frame is represented by the intensity or color of each point in the image.}
% ( Figure \ref{fig:modgd}).
The location of the prominent peak  in the range  corresponds to $[P_{min}, P_{max}]$ in the  {\small{MODGD}}  feature space  is mapped to 
the candidate pitch estimate in the first pass. In the second pass, the estimated pitch component and its harmonics are annihilated from the 
flattened power spectrum. Then the residual signal will be traced for the second frequency component using {\small{MODGD}} analysis. In the post processing phase, pitch grouping followed by
removal of pitch outliers results in the final pitch trajectories. The subsequent sections describe these steps in detail.

% Thus iterative detection and cancellation 
% procedure results in the estimation of partials present in each frame. Multiple  pitches are detected iteratively frame by frame. 

 \subsection{FIR comb filtering}
\begin{figure}[h!]
 \centering
\includegraphics[width=7.5cm,height=4cm]{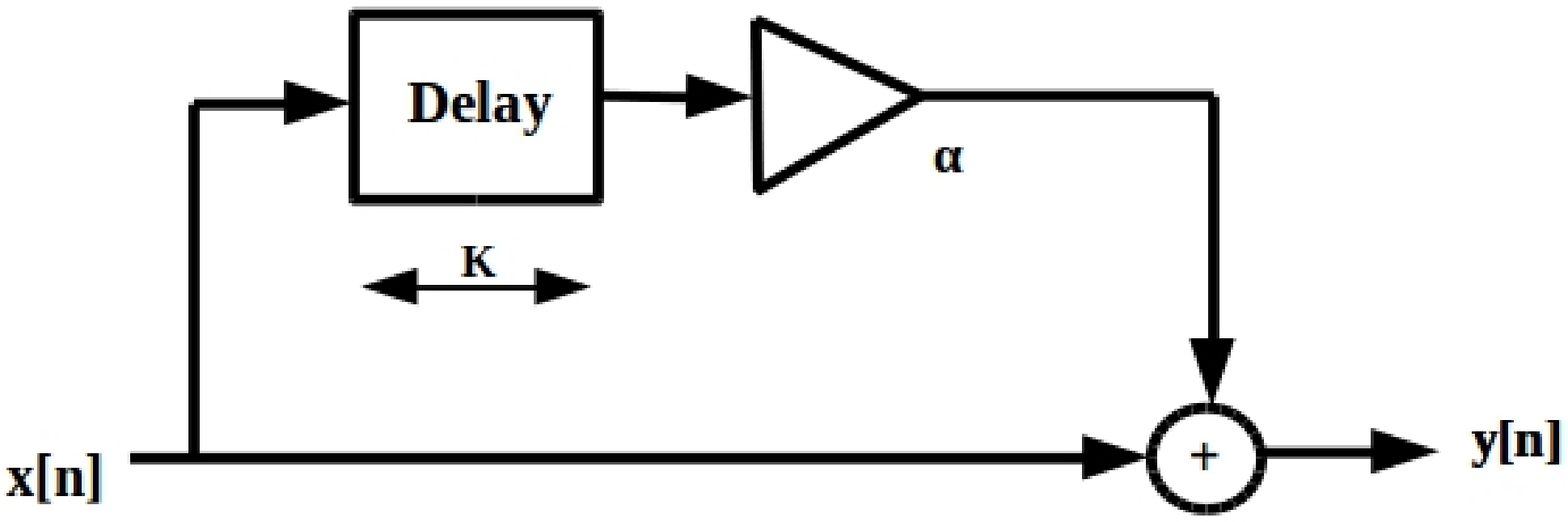}\\
\includegraphics[width=8.5cm,height=8.5cm]{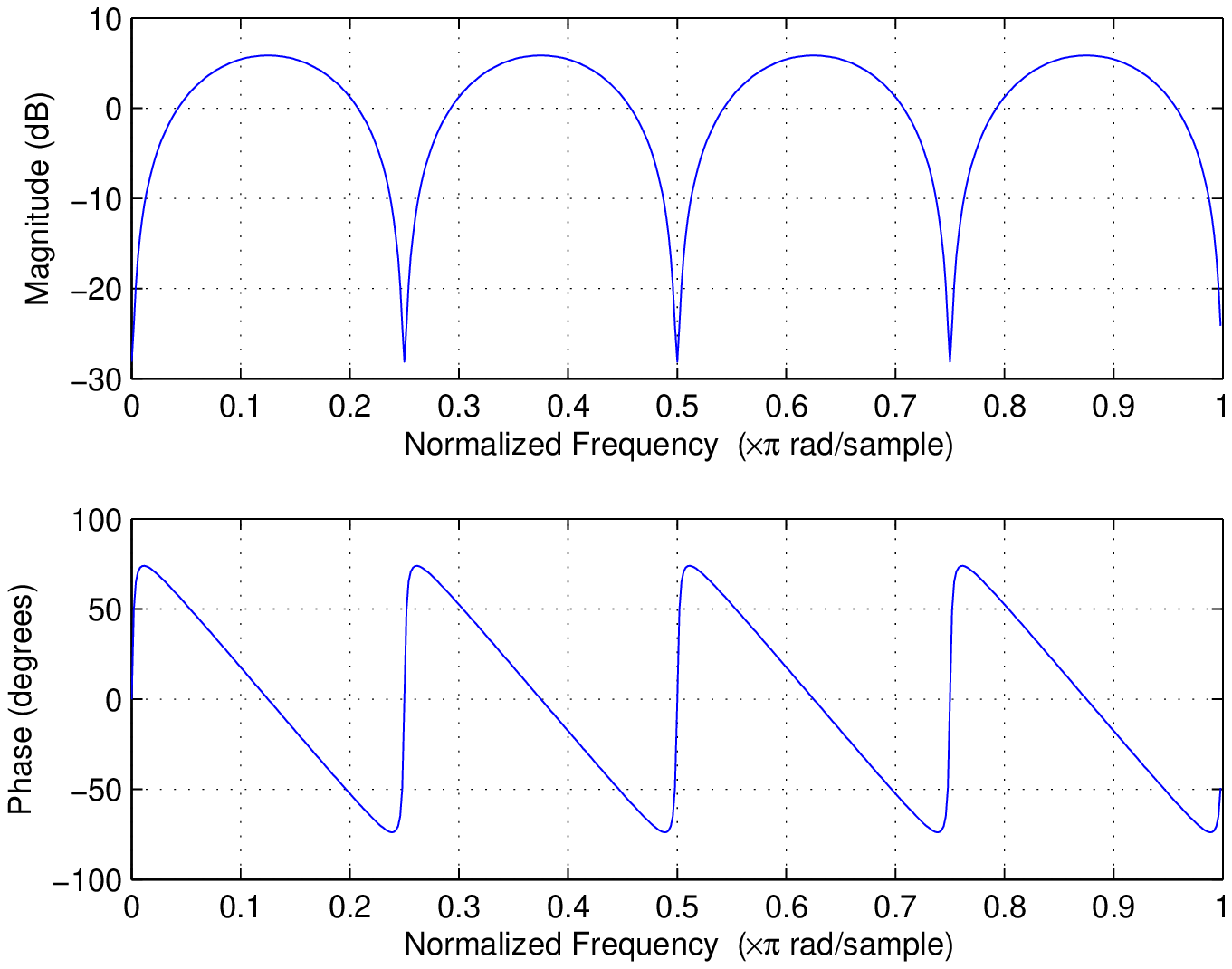}
\caption{\small{Comb filter structure with its magnitude and phase responses }}
\label{fig:combfigure}
\end{figure}

Once the prominent pitch is estimated from the flattened spectrum in the first pass, next we aim at the estimation of the second pitch candidate. 
In the second 
pass, the estimated pitch and its partials
are removed from the flattened spectrum using a comb filter.  
Comb filters are widely used in many speech processing applications such as speech enhancement, pitch detection and speaker recognition \citep{comb2}, \citep{lask}. 
%The proposed algorithm
%uses comb filters to annihilate the estimated pitch and its harmonics from the flattened spectrum in the second pass. 
The FIR  comb filter transfer function is given as :
\begin{equation}
 H(z) =  \frac{Y(z)}{X(z)} = 1+ \alpha z^{-D}
\end{equation}
where $D$ is the pitch period, $\alpha$ a constant  and $X(z)$ , $Y(z)$ represent the $z$-domain representation of input and output respectively.  Magnitude response of the comb filter is 
% \begin{equation}
%  E(z) =(1- \alpha z^{-D})X(z)
% \end{equation}
% 
% 
% 
% 
% 
%  \begin{equation}
%  e[n] =x[n]-\alpha x[n-D]
% \end{equation}

% The comb effect results from phase cancellation and
% reinforcement between the delayed and undelayed
% signal. 
\begin{equation}
\mid H(e^{j\omega}) \mid = \sqrt{(1+ \alpha ^2)+2\alpha \cos(\omega D)}
\end{equation}
The basic structure of the comb filter and its responses are  shown in Figure \ref{fig:combfigure}.
In the proposed approach, comb filter is  used to annihilate the predominant fundamental  frequency component  obtained in the first pass from  a composite flattened power spectrum which constitutes multiple excitations.
\label{multi}
%The proposed method deals with the estimation of individual pitch tracks in a multipitch environment.  

\begin{figure}[h!]
\centering
\includegraphics[width=9cm,height=6cm]{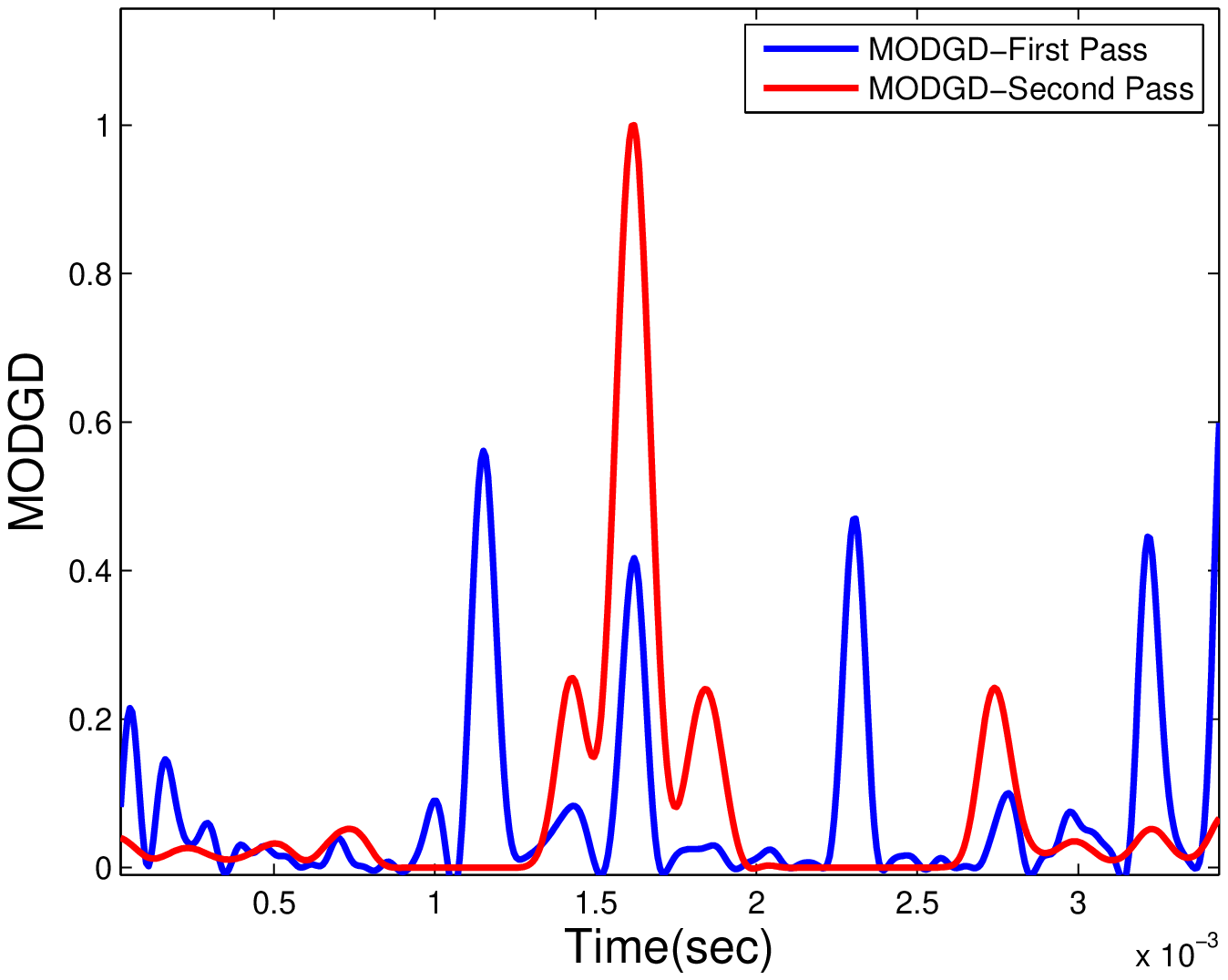}\\
\includegraphics[width=9cm,height=6cm]{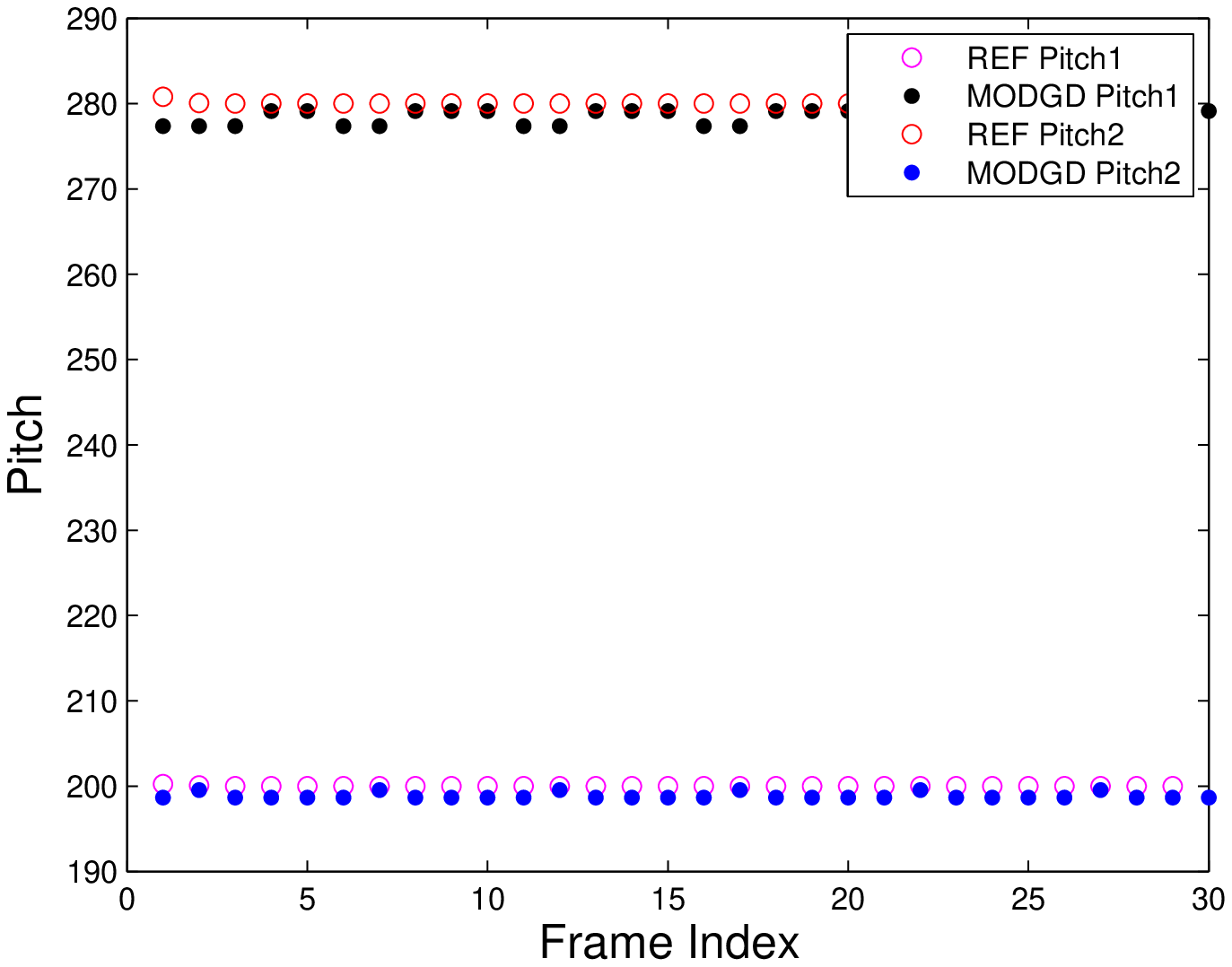}
\caption{\small{ (a) {\small{MODGD}} on the Flattened spectrum for a frame  (b) Pitch extracted for the mixed synthetic speech}}
\label{figsynth}
\end{figure}

\begin{figure}[h!]
\centering
\includegraphics[width=10cm,height=7cm]{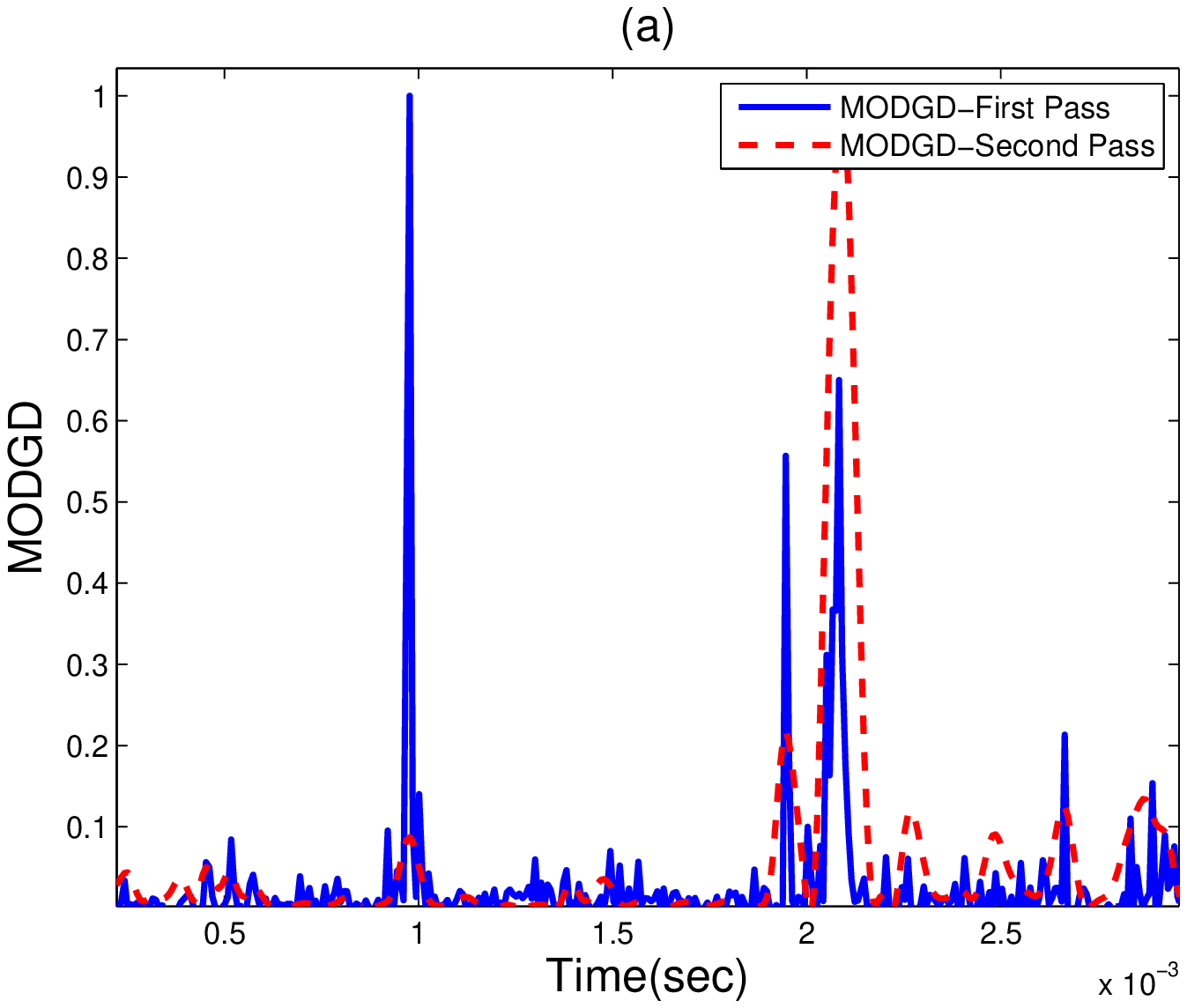} \\
\includegraphics[width=10cm,height=7cm]{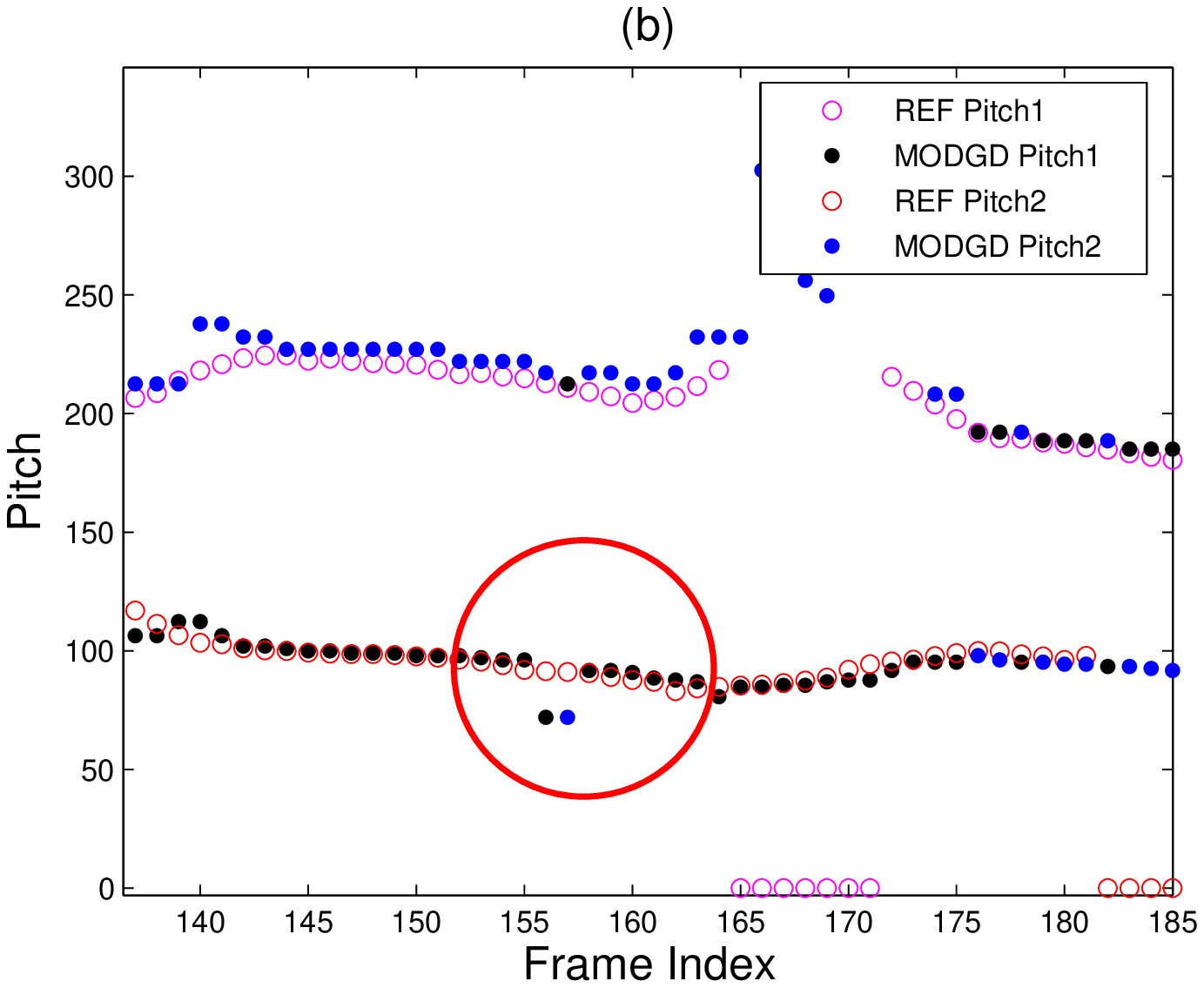}
\caption{{ (a) {\small{MODGD}}} plot for a frame in first pass (blue) and residual spectrum (red) for a real speech mixture, (b) Presence of stray values in the pitch contour are marked in circle }
\label{stray}
\end{figure}

For the instance,  consider a  speech mixture of  two synthetic
speech signals with $f_o$ s 200 Hz and 280 Hz.   
% Let  $\delta_{T_0}[n]$ , $\delta_{T_1}[n]$   be the corresponding excitation components. 
% The composite signal $x[n]$  can be represented by
% \begin{equation}
%  x[n] = \delta_{T_0}[n] +\delta_{T_1}[n]
% \end{equation}
% with 
% % {\small{\begin{equation}
% %  \delta_{T_0}[n] = \sum_{k=-\infty}^{+\infty}\delta[n-kT_0] , \ \ \ \  \delta_{T_1}[n] = \sum_{k=-\infty}^{+\infty}\delta[n-kT_1] 
% % \end{equation}}}
% % 
% % In $z$-domain,
% {\small{\begin{equation}
%  \delta_{T_0}[z] = \sum_{k=-\infty}^{+\infty}\delta[z-kT_0] z^{-{kT_0}} , \ \ \ \  \delta_{T_1}[z] = \sum_{k=-\infty}^{+\infty}\delta[z-kT_1]z^{-{kT_1}} 
% \end{equation}}}
Modified group delay function computed for a synthetic mixture frame is shown in Figure \ref{figsynth}. In Figure \ref{figsynth}(a), blue color plot is the MODGD obtained in the first pass. The peaks in MODGD feature space, correspond to the pitch candidates present in the speech mixture and its integral multiples.
In the first pass, the prominent peak in the {\small{MODGD}} feature space is mapped to first pitch estimate 
followed by the annihilation of it from the the residual spectrum.  The red color contour in  Figure \ref{figsynth}(a) is the computed {\small{MODGD}} for the second pass. 
The individual pitch tracks computed through the aforesaid steps are shown in  Figure \ref{figsynth}(b) along with references.
Similarly, another real audio mixture example is shown in Figure \ref{stray}. Figure \ref{stray}(a) shows the {\small{MODGD}} plot for a real audio frame and  in Figure \ref{stray}(b),  pitch estimates of the audio segment  are shown.  The modified group delay functions obtained in the first pass and in the second pass are illustrated
in the figure. It is obvious from the figure that the  peak corresponds to the predominant pitch  computed in the first pass is annihilated during the second pass.
\subsection{Pitch trajectory estimation by grouping  }
At the end of  the pitch estimation phase, two pitch candidates per frame are computed.  In the pitch grouping stage,  these candidates are grouped into trajectories which comprise continuous, smooth individual tracks. 
A more heuristic approach for grouping is the use of high-low criteria.  Since pitch crossing is not considered, out  of two candidates per frame
high pitch values are grouped into one trajectory and low values to other. 
%A post processing step is done to remove spurious estimate by framing 
\hspace{-0.5cm}

Dynamic programming based pitch grouping can also be employed.  In that case,  the relative
closeness of the distance between peaks in  two consecutive frames is used to compute optimal path.  Transition cost is computed as the absolute difference in distance between the current and previous frame.
The optimal path is selected by minimizing the transition  cost across frames using back tracking approach.  The transition cost $ C_t(c_j/c_{j-1})$  between the  pitch
candidates $ c_j $ and  $c_{j-1}$ of consecutive frames  is given as \citep{veldhius}
%{\small
\begin{equation}
\label{eq:transcost}
     C_t(c_j/c_{j-1}) =  \mid L_j -L_{j-1} \mid
\end{equation}
where $L_j$, $L_{j-1}$ are peak locations in consecutive frames.  
%$l_{max}$
% is the maximum transition distance between peaks, computed from the range  [ $ P_{max} $ , $P_{min}$]. The transition
% cost is  normalized with the  maximum possible transition distance.
The dynamic programming algorithm  finds an optimal pitch sequence $(c_1 ...c_M) $ with  candidates   $ c_1 $ in the first and $c_M $ in the
$M^{th}$ frame  in a block by  minimizing the transition cost function\citep{veldhius}.
% \begin{equation}
% Total\ cost(TC) = Local\ Cost + Transition \ Cost
% \end{equation}
Transition cost  $ TC(c_1 ... c_M)$  of  pitch candidates  $c_1$ to  $c_M$ is
computed by
\begin{equation}
\label{eq:totalcost}
     TC (c_1 ... c_M) = \sum_{j=2}^{j=M}  C_t(c_j/c_{j-1})
\end{equation}
The optimal sequence of pitch markers is determined by back tracking from the
candidate $c_M$ in the $M^{th}$  frame
 in a block to its starting frame.  If the pitch detection algorithm  computes  any spurious candidate, 
 the dynamic programming may result in erroneous pitch tracks.  The proposed algorithm is implemented using the first approach and form pitch contours by ensuring continuity.

% \begin{figure*}[http]
% \centering
% %\includegraphics[width=9cm,height=10cm]{totalsyntheticplot3.eps}
% \includegraphics[width=8cm,height=6cm]{Figures/intermediatePitch.eps}
% %\includegraphics[width=8cm,height=4.5cm]{Figures/synthetic_pitch.eps}\\
% \includegraphics[width=8cm,height=6cm]{Figures/final_pitch1.eps}\\
%  %\includegraphics[width=8cm,height=6cm]{Figures/wupitch2.eps}\\
% % \includegraphics[width=4.25cm,height=4.5cm]{synthetic1_pitch.eps}
% \caption{\small{ (a) Initial pitch estimates for a speech mixture     (b) Final pitch trajectories estimated using the proposed algorithm (c) Pitch trajectories estimated using   WWB  algorithm}}
% \label{finalpitch}
% \end{figure*}
\begin{figure}[h!]
\centering
\includegraphics[width=9cm,height=5cm]{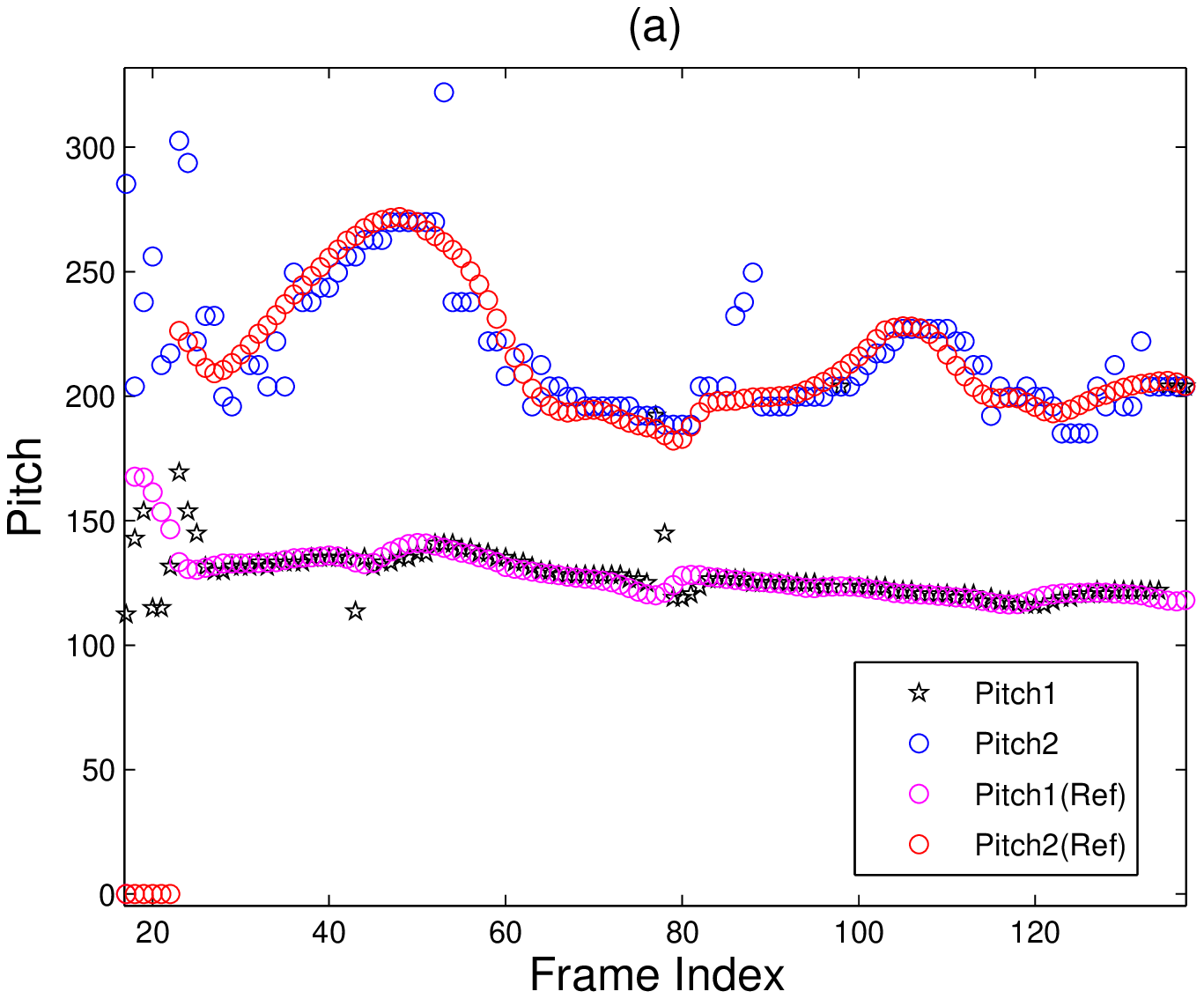}
\includegraphics[width=9cm,height=5cm]{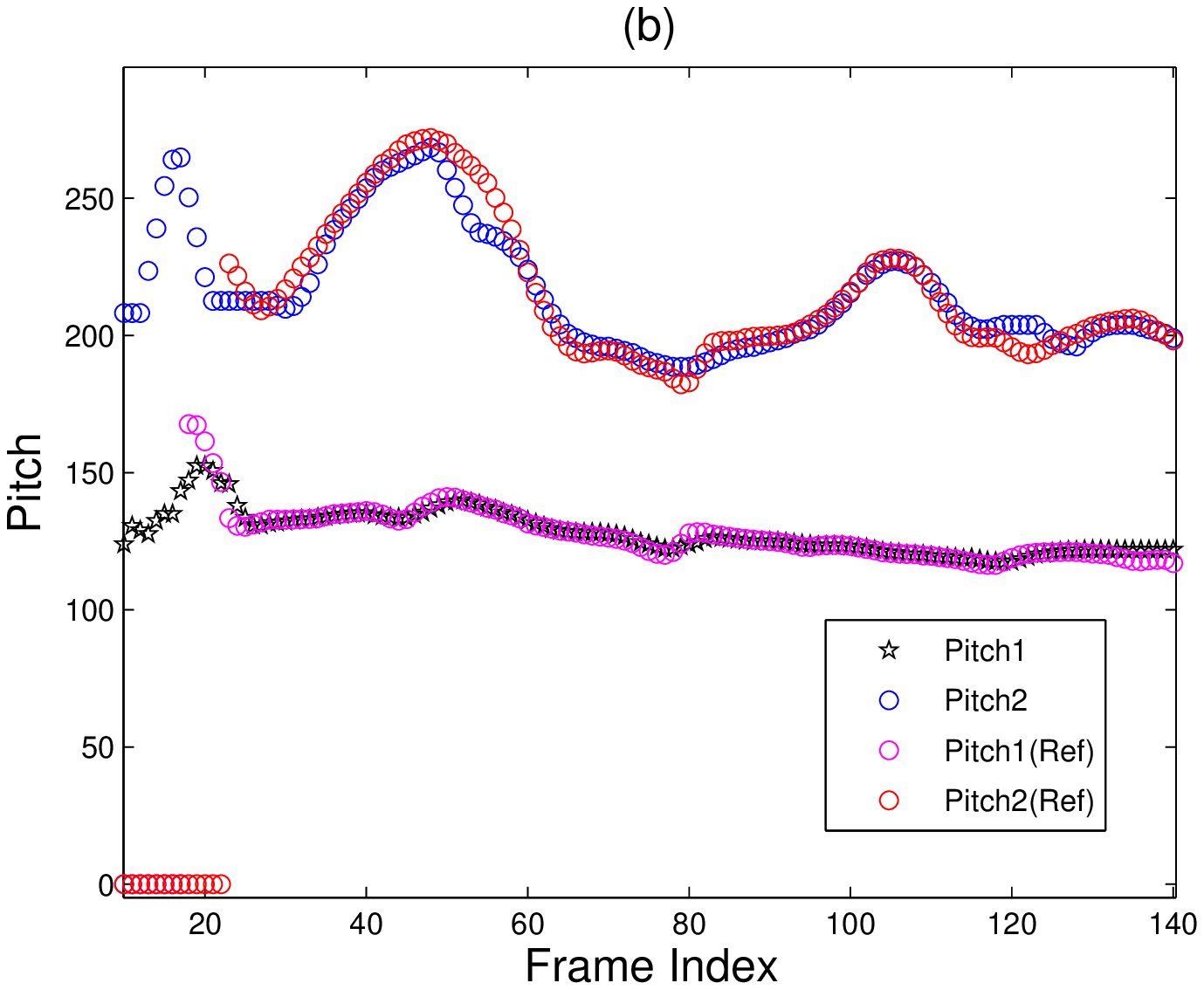}
 \includegraphics[width=9cm,height=5cm]{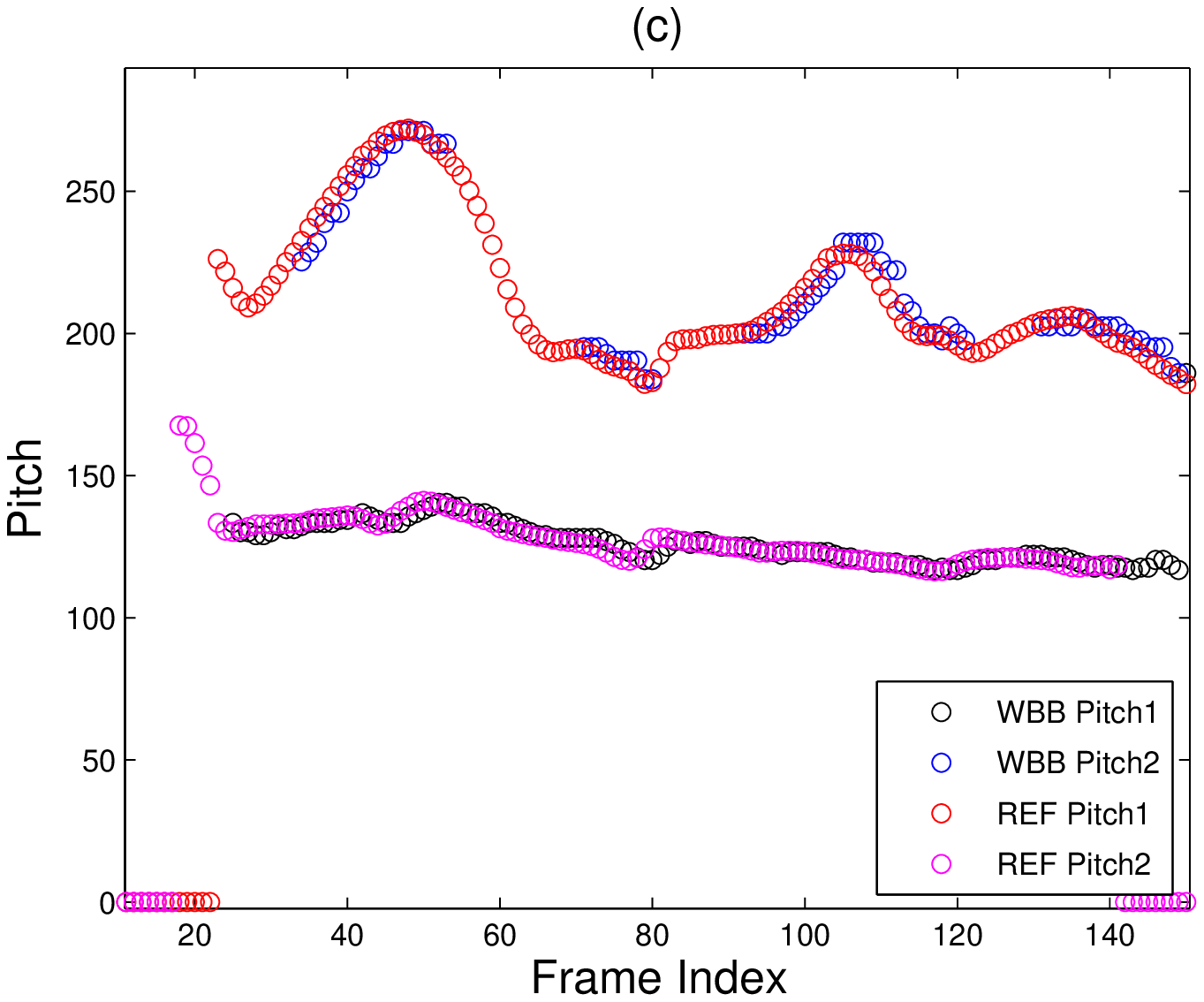}
\caption{\small{ (a) Initial pitch estimates for a speech mixture     (b) Final pitch trajectories estimated using the proposed algorithm (c) Pitch trajectories estimated using   WWB  algorithm}}
\label{finalpitch}
\end{figure}
\subsection{Postprocessing}
The accuracy in pitch estimation is improved by a post processing stage. In this stage, first task is to identify the segments where one or no speaker is present.  A soft threshold on spectral flux is employed to identify these segments. 
%Spectral flux is a measure which shows how quickly the power spectrum of a signal is changing.
The spectral flux is computed as the squared difference between the normalized magnitudes of
the spectral distributions of adjacent frames.
\begin{equation}
F_r =\sum_{k=1}^{N/2} (|(X_r[k]-X_{r-1}[k]|)^2
\end{equation}
where $X_n[k]$ is the magnitude spectrum vector for the $k^{th}$ subband
of frame $n$. Segments which are detected as single speaker frames in the voicing 
 detection stage is processed again for  monopitch estimation using the {\small{MODGD}} algorithm. If the pitch estimated follows a path, estimated sequences are plugged into to the already formed contour 
 by continuity check.

As part of smoothening the curve, stray values will be removed by framing rules to refine the pitch contour, thus minimizing the erroneous pitch estimates.
For example, let  $ f_t $ and  $ f_{t+1} $  be the pitch candidates of consecutive frames in a pitch track after the grouping stage.  If $f_{t+1}$ lies outside the range
[ $f_t$ - $\rho$ , $f_{t}$+ $\rho$ ],  this is treated as a spurious pitch estimate  and will be interpolated  using previous and successive
pitch values \citep{mptracker}. We use linear interpolation for identifying
missing pitch frequencies; however other interpolation techniques
such as cubic or spline interpolation could be used. This simple
but effective technique reduces the pitch error considerably. Note
that missing pitch frequencies should typically not be interpolated
for segments corresponding to 40 msec or longer for typical speech
statistics \citep{mptracker}.   The threshold $\rho$ is set heuristically to 10 Hz. A typical example is shown in  Figure \ref{stray}(b). The circled part indicates the presence of two stray
values in the middle of a continuous curve.
The estimated pitch trajectories for a speech mixture with cross gender pattern is shown in Figure \ref{finalpitch}.  Figure \ref{finalpitch}(a) shows the initial pitch estimates and Figure \ref{finalpitch}(b) shows the individual pitch trajectories after post 
processing. The pitch trajectories estimated using Wu {\it {et al.}} algorithm \citep{wu2} is also shown in  Figure \ref{finalpitch}(c) for the same speech mixture.  
A detail analysis of the results can be seen in the Section \ref{results}.
\subsection{Pitch extraction in noisy and reverberant environment}
The presence of noise and reverberation in speech poses major problems even in monopitch estimation. For noise corrupted speech, both the time-domain
periodicity and spectral-domain periodicity are distorted and hence the conventional pitch estimation fails to certain extent \citep{feng}.
Group delay domain representation of speech makes it relatively immune to noise when
compared to that of the short-time magnitude spectrum \citep{hegde2007b,Gd1}.
%Estimating sinusoids in the presence of noise is well proven in (\citep{Gd1}).
% Consider a signal model consisting of multiple sinusoids (\citep{mustafa}),
% \begin{equation}
%  s[n]=\sum_{k=1}^{K} A_k sin(\omega_kn+\theta_k)
% \end{equation}
% where  $A_k$ is the amplitude, $\omega_k$ is the angular frequency, and $ \theta_k$ is the phase of the $k^{th}$ real sinusoid. ${A_k ; k =
% 1. . . .K}$ and  ${\omega_k ; k = 1. . . .  K}$ are unknown real constants, whereas ${\theta_k ; k = 1. . . . K}$ are assumed to
% be realizations of random variables, distributed uniformly and independently over $[0; 2\pi )$.  K is the number
% of sinusoids, and $N$ is the sample size. It is well known that $s[n]$ obeys the following autoregressive difference equation
% 
% \begin{equation}
%  B(q^{-1})s[n] =0
% \end{equation}
%  where $ q^{-1}$ denotes the unit delay operator, $q^{-1}s[n] =s[n-1] $ and  $B(q^{-1})$ is a polynomial of degree $2K$ given by 
%  
%  {\small{\begin{equation}
%   B(q^{-1}) = 1+b_1 q^{-1}+....+b_{2K} q^{-2K} =\prod_{m=1}^{K}(1-2\cos \omega_m q^{-1} + q^{-2})
%  \end{equation}}}
 For instance, consider the noisy signal $x[n]$ as the output of the autoregressive process $s[n]$, corrupted with  Gaussian noise $ \omega[n]$, 
 i.e
 \begin{equation}
 x[n]=s[n]+\omega[n]
  \label{eqn1}
\end{equation}
Group delay analysis of an autoregressive process in a noisy environment is given in \citep{Gd1}.  $Z$ transform of $s[n]$,
ignoring the effects of truncation of the response of an all-pole system is given as  
 
 \begin{equation}
 S(z) = \frac{GE(z)}{A(z)}
 \label{eqn2}
\end{equation}
  
 where $E(z)$ is the z transform of the excitation sequence
$e[n]$ and $G / A ( z ) $is the z transform of the all-pole system
corresponding to the autoregressive process.  From Equations (\ref{eqn1}) and (\ref{eqn2})
\begin{equation}
 X(z)  = \frac{GE(z) + W(z)A(z)}{A(z)} = \frac{V(z)}{A(z)}
\end{equation}
% ,where
% \begin{equation}
%  V(z) = GE(z) + W(z)A(z).
% \end{equation}
In group delay domain,
% The group delay function of $X ( z )$ in terms
% of the delay functions of $V(z)$ and $A(z)$ is given by
 \begin{equation}
 \tau_X(\omega) = \tau_V({\omega})- \tau_A(\omega)    
\end{equation}

\begin{figure}[h!]
\centering
\includegraphics[width=8cm,height=5.5cm]{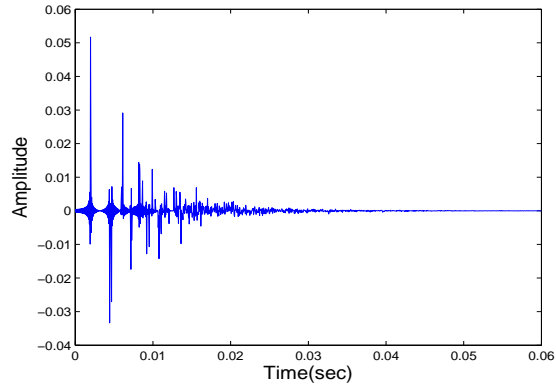}
\caption{\small{ Room Impulse Response }}
\label{fig_sing_speaker}
\end{figure}

As explained in \citep{Gd1}, the noise spikes in $\tau_X({\omega})$ can be suppressed by  multiplying  with the estimated zero spectrum. This 
results in an estimate  of  -{$\tau_A({\omega})$} which  corresponds to  the spectral component in the composite signal. Thus group delay based approach is 
very effective in analyzing frequency components of a composite signal in the presence of noise. 
Room reverberation adversely affect the characteristics
of pitch  and thus makes the task of pitch determination more challenging.
It causes degradation of the excitation signal due to the received speech signal because of the involvement 
of another filter which characterizes the room acoustics \citep{jinWang2}.
In reverberant environments, the speech signal that reaches the microphone is superimposed
with multiple reflected versions of the original speech signal. These superpositions can be modeled
by the convolution of the room impulse response (RIR), that accounts for individual reflection delays,
with the original speech signal \citep{allen}.
% The acoustic pressure pattern induced at a particular point in a
% room in response to a pressure impulse of unity magnitude at another point in the room is termed as RIR(Room Impulse Response).
Mathematically, the reverberant speech  $r[n]$ is obtained as the convolution of speech signal $s[n]$ and room impulse response $h[n]$ \citep{samuel}.
%  Let denote the original speech signal and  the room impulse response respectively.
%  Then the reverberant speech given by .
\begin{equation}
        r[n] = s[n] \ast h[n]
\end{equation}
% Polack developed a time-domain model complementing Schroeder's frequency-domain model  \citep{jean} .
Room impulse response is described as one realization of a non-stationary
stochastic process in  Schroeder's frequency-domain model \citep{jean1} as 
\begin{equation}
h[n] = b[n] e ^{-\delta n} ,for ~~n \geqq 0
\end{equation}
where $b[n]$ is a centered stationary Gaussian noise, and $\delta$ is related to the reverberation time $T_r$. A typical
room impulse response used for the experiment is shown in Figure \ref{fig_sing_speaker}. The proposed algorithm is also analysed in a reverberative condition using simulated impulse
reponse.

% The random noise $b(t)$ is characterized by its power spectral density, denoted $P(f)$
% The effect of reverberation on the short-time Fourier transform (STFT)
% of the speech signal $s(t)$ can be represented as
% \begin{equation}
% \label{eqn0}
%         R( \omega) = S( \omega)H(\omega)
% \end{equation}
% 
% where $S( \omega) $ and $ R( \omega)$ are the STFT’s of $s(t)$ and $r(t)$ respectively. $H( \omega) $ denotes the STFT of the room iimpumpulse response $h(t)$.
% 
% Let  $R_{cep}(\omega)$ represents the cepstrally smoothed version of $ R(\omega)$, then 
% \begin{equation}
% \label{eqn1}
%     E_{rev}( \omega)  = \frac{ R( \omega)}{R_{cep}(\omega)} 
% \end{equation}
%  Substituting  equation \ref{eqn0} in equation \ref{eqn1} 
% \begin{equation}
% \label{eqn2}
%     E_{rev}( \omega)  = \frac{S(\omega)H( \omega)} {R_{cep}(\omega)}
% \end{equation}
% 
% \begin{equation}
% \label{eqn2}
%     E_{rev}(\omega)  = \frac{S(\omega)}{R_{cep}(\omega)} H( \omega)
% \end{equation}
% In the group delay domain,
% 
% \begin{equation}
% \label{eqn4}
%     \tau_{rev}(k)  = \tau_{e}(k)  + \tau_h(k) 
%  \end{equation}
% 
% where $\tau_{rev}(k)$ , $\tau_{e}(k)$ and $ \tau_h(k)$ are the group delays of excitation component in the reverberated signal, 
% excitation of speech signal and the room impulse response respectively.
%  The Figure \ref{fig_sing_speaker}(b) shows the location of the peaks in {\small{MODGD}} feature space in clean and reverberant condition for a speech frame. The location of the peak is preserved in a reverberative environment.
% Pitch estimated using the modified group delay algorithm for a single speaker utterance  is shown in Figure \ref{fig_sing_speaker}(c) for $T_{60}$ = 0.2 sec.
\begin{table}[h!]
\label{category}
\centering 
\caption{Category of mixtures for Dataset:1 and 2 } 
\begin{center}
\scalebox{1.0}{
\begin{tabular}{ p{1cm}  p{8cm} }
\hline
\textbf{Category}   ~ & ~~~~~~\textbf{Speech data}  \\ \hline
1 & Male/Female, Female/Female, Male/Male\\  
2 & Male/Female, babble noise \\ 
3 & Male/Female, white noise  \\
4 & Male/Female  with reverberation \\
 \hline
\end{tabular}}
\end{center}
\label{category}
\end{table}

\section {Evaluation }
\label{dataset}
\subsection{Evaluation  data set}
In the proposed work, focus is given to the multipitch estimation of speech mixture with two speakers. The performance of the proposed algorithm was evaluated using following  datasets:- 
\begin{itemize}
 \item  Dataset-1:  The dataset consists of 40 audio files obtained by mixing a subset of utterances from 
the Pitch Tracking Database of  Graz university of Technology (PTDB-TUG)  \citep{ pirker}, Childers database  and a few audio
samples from $ Simple^4 All$  speech corpora \citep{sampleforall}.   The PTDB-TUG consists of audio recordings with phonetically rich sentences
from  TIMIT corpus.  The TIMIT corpus consists of dialect sentences  (labeled as {\it{sa}}),  phonetically-compact sentences (labeled as {\it{sx}}),  and phonetically-diverse sentences(labeled as  {\it {si}}). 
$ Simple^4 All $ corpora consists of audio samples from different languages. 
\item  Dataset-2:  GRID \citep{grid1} is a large multitalker audiovisual sentence corpus to support joint computational-behavioral studies in speech perception. The corpus
consists of high-quality audio and video  recordings of 1000 sentences spoken by each of 34 talkers (18 male, 16 female).  A subset of 40 audio files are used  for generating mixtures for the evaluation.
\end{itemize}

In the experiments, each audio mixture is processed using a hamming window of  frame length of 30 ms and hop size of 10 ms.
As shown in Table \ref{category}, the interferences are classified into four categories by considering clean and noise conditions.  The data set contains audio files of  cross gender (male/female) and same gender (female/female, male/male) patterns .
The test was conducted mainly on 0 dB 
target-to-masker ratio (TMR) which is considered the most 
difficult situation in co-channel speech segregation problem as 
both talkers equally mask each other.   In category 2 and 3, speech is obtained by mixing the category 1 speech data with babble noise (5 dB SNR) and white noise (10 dB SNR).  Category 4 interferences comprising
of simulated reverberant speech utterances. The performance is also evaluated with speech mixture generated by clean voices of cross gender pattern with +3dB and -3dB Target to Masker Ratio(TMR).

\begin{table*}[!htbp]
\centering
\caption{Comparison of Accuracy (Dataset:1)}
\scalebox{0.725}{
\begin{tabular}{*9c}
\toprule
Category &  \multicolumn{4}{c}{${ Accuracy_{20}(in \%)}$}\hspace{1cm} & \multicolumn{4}{c}{$ {Accuracy_{10}(in \%) }$}\\
\midrule
      {	}                      & {\small{MODGD}}      & WWB      &JIN     \hspace{1cm}     &{\small{MODGD}}    &WWB          &JIN            \\
\midrule
Male-Female                    & 88.52                & 77.95    &81.99   \hspace{1cm}     & 84.58          &76.71         &81.04             \\
Female-Female                  & 85.28                & 64.54    &72.00   \hspace{1cm}     & 75.02          &60.78         &70.31              \\
Male-Male                      & 80.53                & 66.24    &72.41   \hspace{1cm}     & 73.58          &66.01         &70.01               \\
Male-Female,Babble noise       & 84.68                & 60.56    &76.90   \hspace{1cm}     & 72.12          &60.17         &73.43               \\
Male-Female,White noise        & 78.04                & 63.11    &77.59   \hspace{1cm}     & 73.13          &62.97         &76.21                \\
Male-Female with reverberation & 74.01                & 73.08    &81.15   \hspace{1cm}     & 63.05          &72.70         &80.28             \\
\bottomrule
\end{tabular}}
\label{comparison3}
\end{table*}
% \begin{figure}[h!]
% %\centering
% % \includegraphics[width=7.75cm,height=5.00cm]{modgdplotReal.eps}\\
% \centering
% \includegraphics[width=6.75cm,height=5.00cm]{Figures/modgdgram1.eps}\\
% \caption{{Modgdgram for a  speech segment}}
% \label{fig:modgd}
% \end{figure}
Reverberant speech is  generated using simulated room acoustics 
 using a MATLAB implementation \citep{lehman}
 %[16] 
 from the image model
\citep{allen}. The model produces the room impulse response (RIR)
when fed with room dimensions, wall reflection coefficients
and physical locations corresponding to sound sources and
the microphone. The simulation is done for reverberation time $T_{60}$ = 200ms.
\begin{table*}[!htbp]
\centering
\caption{Comparison of Accuracy (Dataset:2)}
\scalebox{0.725}{
\begin{tabular}{*9c}
\toprule
Category &\multicolumn{3}{c}{${ Accuracy_{20}(in \%)}$}  \hspace{1cm} & \multicolumn{3}{c}{$ {Accuracy_{10}(in \%) }$}\\
\midrule
      {	}                      & {\small{MODGD}}    & WWB   & JIN  \hspace{1cm} &{\small{MODGD}}   &WWB        & JIN             \\
\midrule
Male-Female                    & 87.88        & 79.58   &79.95  \hspace{1cm}     & 82.65         &78.92      &78.95     \\
Female-Female                  & 78.99        & 77.17   &77.30   \hspace{1cm}    & 74.86         &76.74      &76.81  \\
Male-Male                      & 74.39        & 50.92   &73.50   \hspace{1cm}    & 65.76         &50.84      &73.08      \\
Male-Female,Babble noise       & 77.40        & 57.29   &74.09   \hspace{1cm}    & 70.00         &56.25      &72.74    \\
Male-Female,White noise        & 65.68        &66.96    &72.66   \hspace{1cm}    & 58.73         &66.34      &71.64       \\
Male-Female with reverberation & 73.30        &64.44    &79.00   \hspace{1cm}    &68.00          &64.01      &78.38       \\
\bottomrule  
\end{tabular}}
\label{comparison3b}
\end{table*}

\subsection{Evaluation  metrics}
The performance is evaluated only for voiced frames.
The reference frequency of an unvoiced frame is considered as 0 Hz.  %Experiments are carried out to validate our algorithm by 
%evaluating the accuracy of $f_0$ detection in comparison with 
%well-known cepstrum method and   WWB  algorithm.
To evaluate the performance of our algorithm, requires a  reference pitch contour corresponding to the true individual pitch. We computed the reference 
pitch of clean speech using Wavesurfer \citep{jin4}. The guidelines for evaluating the performance of monopitch estimation can be seen in \citep{rabinerPitch}.
Since there are no generally accepted guidelines  for the performance evaluation in the case of multipitch tracking, we extended 
the guidelines of single pitch tracking.
The performance is quantitatively assessed by measuring two types of metrics:
accuracy and  standard deviation of the fine pitch errors  $ E_{fs} $. \\
%pitch estimation errors: Gross detection error $ E_{gross}$, Standard Deviation of the Fine Pitch Errors $ E_{fine} $
The metrics are defined as follows,

\begin{itemize}
 \item \textbf{Accuracy}:  $Accuracy_{10}$ and $Accuracy_{20}$ correspond to the  percentage of frames at which pitch deviation is less than 10\%  and 20\%   with respect to the reference respectively.   
A gross error occurs if the detected pitch is not within the specified threshold with respect to the  reference pitch. 

\vspace{0.3cm}

 \item\textbf {Standard deviation of  the fine pitch errors  ($ E_{fs} $)}: The standard deviation of the fine pitch error is a 
measure of the accuracy of the pitch detection during voiced intervals.
 The standard deviation of the pitch detection $\sigma_e$ is 
given as: 
{\small
\begin{equation}
  \sigma_e = \sqrt(\frac{1}{N} \sum\limits(p_s - p'_s)^2 -e^2
\end{equation}}
where $p_s$ is the standard pitch, $p'_s$ is the detected pitch, $N$ 
is the number of correct pitch frames and $e$ is the mean 
of the fine pitch error. $e$ is given as:

%{\small
\begin{equation}
  e = \frac {1}{N} \sum\limits (p_s - p'_s)
\end{equation}
\end{itemize}

\section{Analysis and Discussions}
\label{results}
The performance of the proposed algorithm was evaluated primarily on  speech mixtures, speaking simultaneously with equal average power.  Three patterns, same gender (M/M, F/F) and cross gender (M/F) are 
considered for the evaluation. In addition to the clean speech condition, the performance is also evaluated
in  noisy and reverberant conditions.  WWB  algorithm \citep{wu2} and Jin {\it {et al.}} algorithm \citep{jin3}  are used for the objective 
comparison in performance evaluation.   
The algorithm of Wu, Wang, and Brown is referred to as the WWB algorithm. 
WWB algorithm  integrates a channel-peak selection 
method and Hidden Markov Model (HMM) for forming continuous pitch 
tracks. WWB framework computes final pitch estimates in three stages: auditory front-end-processing, pitch statistical modelling and HMM tracking. Jin {\it {et al.}}
algorithm, designed specially to tackle reverberant noises is similar to WWB algorithm but different in channel selection  and pitch scoring strategy. An auditory
front-end and a new channel selection method are utilized to extract periodicity features in Jin {\it {et al.}} algorithm.  
In  \citep{wu2,jin3}, half of the corpus is used to estimate the model parameters and thus supervisory in nature. Another important fact
about the experiments reported in \citep{wu2} is that  they are focused on speech mixtures with one dominating speaker. Both this algorithms report considerable amount of transition errors, in which   pitch estimates of speaker-1
are misclassified as the pitch estimates of speaker-2.  For a fair 
comparison, the  WWB and  Jin {\it {et al.}} algorithm outputs are grouped in the post processing stage to ensure no transition error is occurred \citep{jin3}.

The grouping is done using a similar approach proposed in  \citep{mptracker}.
We consider each track as a cluster of data and the
mean of each cluster as representative of that cluster. Let $g \{q \} =
\{\omega^t , . . . , \omega^{(t+p)} , . . . , \omega^{(t+P-1)} \} $  be the $q^{th}$ track (or equivalently cluster) with length $P$ . Then the mean of the $q^{th}$ cluster is defined as
$M^q$ = $ \frac{1}{P}(\sum_{p= 0}^{ p-1} \omega^{(t+p)})$.
For the two-speaker case, pitch tracks are
classified into two groups (I and II), one belonging to each speaker.
To do this,  mean of the first segment in each track is computed as  $M^{1*}$ and $M^{2*}$ respectively. Then successive segments are grouped into
one of the tracks by assessing the closeness of $M^q$  with $M^{1*}$ and $M^{2*}$ by fixing a threshold  $k$.
% track and segmentsits mean is found the longest track $q^∗$ is identified and its mean, $M^*$ ,
% is compared with the means of other tracks. Those tracks whose
% means satisfy $\mid M^q - M* \mid > k$ are classified into group II. The remaining tracks are classified into group I including the longest track.
%In this way, we obtain two groups, each containing the pitch contours of one speaker.

%The underlying speech signals are mixed with the target-to-interference ratio (TIR) of 0, 3, 6, 9, 12, 15, and 18 dB.
The results obtained through quantitative evaluation
are listed in  Tables \ref{comparison3} - \ref{comparison7}. 
Table \ref{comparison3} and  \ref{comparison3b} compare 
the  pitch accuracies with 20\% and 10\% tolerance for dataset-1 and dataset-2 respectively. 
The results can be used to analyse the performance of the proposed system in clean  and noisy conditions with same/cross gender speech mixtures.
In clean conditions, the proposed group delay based system outperforms the other two systems. Jin {\it {et al}} algorithm and MODGD algorithm show a 
neck to neck performance giving slight advantage to MODGD system. Another important point we noticed in the 
experiment is that WWB algorithm fails to pick one of the pitch estimates in many frames. 
The proposed method reports accuracies of   84.58\% and  
82.65\% within  10\% tolerance  for dataset 1 and dataset 2 respectively in clean mixtures. In noisy conditions,
Jin {\it{et al}}, algorithm shows good performance especially in reverberant conditions. It is worth in  noting that, in babble and white noise conditions,
MODGD system is at par with the Jin {\it{et al}} algorithm  and also shows a superior performance over WWB algorithm.
In same gender mixture
patterns, if the pitch values are too close, the performance  of the proposed algorithm is affected due to the filtering operation.
% It is worth in  noting that  in the presence of noise 
% condition {\small{MODGD}} based algorithm outperforms  WWB  algorithm.
In the proposed group delay based system, both noise and source introduce zeroes that are close to the unit circle in the $z$ domain \citep{hemathesis,rajeshthesis}.  The fundamental difference is 
that source zeroes are periodic while noise zeroes are aperiodic.  This is the primary reason why the proposed algorithm  extracts pitch in the noisy
environment.  Even though in the anechoic 
condition, the proposed system and WWB algorithm yielding competitive performance, in reverberant environment, the performance
of the proposed system is poor as compared to WWB algorithm.
Table \ref{comparison4} and \ref{comparison4b} 
compare the standard deviation of fine pitch error($ E_{fs} $) for dataset-1 and dataset-2 respectively. 
WWB  algorithm and Jin {\it {et al.}} algorithms give $ E_{fs} $ in the range 2-4 Hz across the entire interference categories while the proposed algorithm 
reports slightly high $ E_{fs} $ in the range 3- 5.5Hz. 
% As  given  in Table \ref{comparison4} and \ref{comparison4b}, performance of the proposed approach
% is at par with that of  WWB algorithm  for the metric $ E_{fs} $.
Finally, we have done analysis on varying the Target to Masker Ratio (TMR) in the clean conditions. 
The results are tabulated in Table \ref{comparison6} and Table \ref{comparison7}. The analysis shows a similar trend in the performance of both 
MODGD algorithm and Jin {\it et al} algorithm. 
A considerable variation in accuracy is reported in the case of WWB algorithm as TMR varies from -3dB to 3dB, but 
for the other two algorithms, the variation is not that much significant. When it comes to
fine pitch $ E_{fs} $, the variation over different TMR is minimal as compared to  equal TMR situation.
\begin{table}[h!]
\centering
\caption{Comparison of  $ E_{fs} $ corresponding to $ {Accuracy_{10}}$  (in Hz) (Dataset:1) }
\scalebox{0.8}{
\begin{tabular}{*8c}
\toprule
Category  & \multicolumn{2}{c}{~~~~~~~$~~~~~~~~ E_{fs} $}\\
\midrule
      {	}                     &{\small{MODGD}}    &WWB       & JIN      \\
\midrule
Male-Female                        & 4.80          &1.81      &3.12      \\
Female-Female                      & 5.43            &2.12       &4.05       \\
Male-Male                          & 4.82            &1.52       &2.75      \\
Male-Female,Babble noise           & 5.40            &2.17       &3.47     \\
Male-Female, White noise           & 5.47            &2.01       &3.82      \\
Male-Female  with reverberation    & 4.98           &2.87       &3.76    \\          
\bottomrule
\end{tabular}}
\label{comparison4}
\end{table}
 \begin{table}[h!]
\centering
\caption{Comparison of  $ E_{fs} $ corresponding to $ {Accuracy_{10}}$ (in Hz) (Dataset:2) }
\scalebox{0.8}{
\begin{tabular}{*7c} 
\toprule
Category  & \multicolumn{2}{c}{~~~~~~~$ ~~~~E_{fs} $}\\
\midrule
      {	}                     &{\small{MODGD}}           &WWB    &JIN\\
\midrule
Male-Female                        & 3.48           &1.59     &2.37        \\
Female-Female                      & 3.77          &2.17      &3.28      \\
Male-Male                          & 3.65           &1.43      &2.45       \\
Male-Female, Babble noise          & 3.86          &1.61      &2.64           \\
Male-Female, White noise           & 4.27           &1.91       &3.1           \\
Male-Female  with reverberation    & 4.34          &2.55       &3.00           \\          
\bottomrule
\end{tabular}}
\label{comparison4b}
\end{table}

\begin{table*}[!htbp]
\centering
\caption{Comparison of accuracy in various TMR:Database:1}
\scalebox{0.75}{
\begin{tabular}{*9c}
\toprule
Category   & \multicolumn{2}{c}{WWB}  &\multicolumn{2}{c}{JIN}  &\multicolumn{2}{c}{MODGD}                      \\
\midrule
      {	}        \hspace{0.4cm}                    \hspace{0.4cm} &${ Accuracy_{10}(in \%)}$ & $ E_{fs} $     \hspace{0.4cm}   & ${ Accuracy_{10}(in \%)}$   & $ E_{fs} $   \hspace{0.4cm} &${ Accuracy_{10}(in \%)}$  & $ E_{fs} $           \\
\midrule
Male-Female     0dB                                \hspace{0.4cm}    &76.71     & 1.81                                \hspace{0.4cm}   & 81.04              & 3.12     \hspace{0.4cm}    & 84.58     & 4.80            \\
Male-Female   -3dB                             \hspace{0.4cm}    & 72.90    & 1.80                               \hspace{0.4cm}   & 79.65              & 3.74      \hspace{0.4cm}    & 83.32    & 4.93            \\
Male-Female      +3dB                                 \hspace{0.4cm}   & 79.40     & 1.71                                 \hspace{0.4cm}   & 82.21              & 3.42     \hspace{0.4cm}    & 85.52     & 4.85             \\
%Male-Female,Babble noise       & 84.68            & 68.01     \hspace{1cm}     & 70.18     & 59.54  \hspace{1cm}    & 87.00           & 72.37       \hspace{1cm}    & 81.58     & 71.23              \\
%Male-Female,White noise        & 78.04            & 69.90       \hspace{1cm}    & 71.89     & 62.27   \hspace{1cm}& 87.00              & 72.37       \hspace{1cm}    & 81.58     & 71.23                  \\
%Male-Female with reverberation & 74.01            & 70.51       \hspace{1cm}    &63.05      & 61.68   \hspace{1cm}  & 87.00             & 72.37       \hspace{1cm}    & 81.58     & 71.23               \\
\bottomrule
\end{tabular}}
\label{comparison6}
\end{table*}

\begin{table*}[!htbp]
\centering
\caption{Comparison of accuracy in  various TMR-Database:2}
\scalebox{0.75}{
\begin{tabular}{*9c}
\toprule
Category   & \multicolumn{2}{c}{WBB}  &\multicolumn{2}{c}{JIN}  &\multicolumn{2}{c}{MODGD}                      \\
\midrule
      {	}        \hspace{0.4cm}                    \hspace{0.4cm} &${ Accuracy_{10}(in \%)}$ & $ E_{fs} $     \hspace{0.4cm}   & ${ Accuracy_{10}(in \%)}$   & $ E_{fs} $   \hspace{0.4cm} &${ Accuracy_{10}(in \%)}$  & $ E_{fs} $           \\
\midrule
Male-Female     0dB                                \hspace{0.4cm}    & 78.92     & 1.59                                 \hspace{0.4cm}   & 78.95              & 2.37                 \hspace{0.4cm}    & 82.65     & 3.48           \\
Male-Female   -3dB                             \hspace{0.4cm}    & 71.02     & 1.73                                \hspace{0.4cm}   & 76.71              & 2.05             \hspace{0.4cm}    & 82.72     & 3.46            \\
Male-Female       +3dB                                 \hspace{0.4cm}   & 74.37     & 1.61                                 \hspace{0.4cm}   & 77.61              & 2.10                \hspace{0.4cm}    & 82.26     & 3.49             \\
%Male-Female,Babble noise       & 84.68            & 68.01     \hspace{1cm}     & 70.18     & 59.54  \hspace{1cm}    & 87.00           & 72.37       \hspace{1cm}    & 81.58     & 71.23              \\
%Male-Female,White noise        & 78.04            & 69.90       \hspace{1cm}    & 71.89     & 62.27   \hspace{1cm}& 87.00              & 72.37       \hspace{1cm}    & 81.58     & 71.23                  \\
%Male-Female with reverberation & 74.01            & 70.51       \hspace{1cm}    &63.05      & 61.68   \hspace{1cm}  & 87.00             & 72.37       \hspace{1cm}    & 81.58     & 71.23               \\
\bottomrule
\end{tabular}}
\label{comparison7}
\end{table*}
\section{Source-MODGD cepstral features in estimating number of speakers.}
\label{smcc}
 In literature, many multipitch estimation algorithms start from the estimation of number of speakers. In  \citep{kameoka2}, a frame 
 independent process is described, that gives good estimates of the
number of speakers and $f_o$s with a single-frame-processing.
The algorithm explained in \citep{kameoka3}  detects number of concurrent speakers based on maximum likelihood estimation of the model parameters
using EM algorithm and  information criterion. In the work proposed by S.Vishnubhotla {\it{et al.}}\citep{vishnu}, the temporal
evolution of the 2-D AMDF is used to estimate the number of
speakers present in periodic regions.
\begin{figure}[h!]
\centering
\includegraphics[width=8cm,height=6cm]{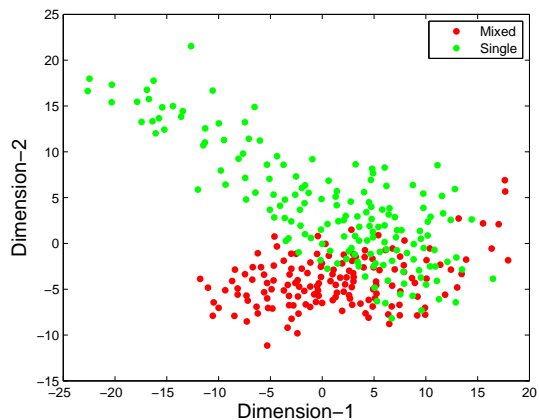}
\caption{\small{ Two dimensional visualization of SMCC features for a single speaker and a speech mixture
using Sammon mapping }}
\label{sammon}
\end{figure}
The proposed method can also be extended to speech mixture with more than two speaker, if the information regarding the number of speakers is available.
The  iterative cancellation steps are determined by the number of speakers present in the mixture. Our experiments show that a variant of group delay feature;
SMCC (Source-MODGD Cepstral features) derived from the flattened spectrum can be efficiently utilized to estimate the number of speakers.
Since the modified group delay function behaves like a squared magnitude response \citep{hamsadhana}, 
homomorphic processing approach can be employed to convert modified group delay spectra to meaningful features.
% Since it is shown that modified group delay behaves similar to square of magnitude spectrum,{\small{MODGD}} feature space can be analysed 
% using filterbank analysis.
The filter bank analysis on {\small{MODGD}} of the flattened power spectrum followed by {\small{DCT}} results in the proposed {\small {SMCC}} feature. \\
 Steps to compute Source-MODGD Cepstral Coefficient features are summarized  below \\
  \begin{itemize}
  \item  Frame blocking the speech signal at a frame size of 20 ms  and frame shift of 10 ms. A hamming window is applied on each frame.
  \item  Speech power spectrum is flattened using  the spectral envelop obtained by cepstral smoothing to annihilate the system characteristics.
   \item Apply {\small{MODGD}} algorithm on the flattened power spectrum to compute modified group delay function of the smoothed spectrum.
  \item  Apply  filter-bank on modified group delay $\tau_m(k)$ to get the Filter Bank Energies (FBEs). 
  \item  Compute {\small{DCT}} of log FBEs to get the {\small {SMCC}} feature vectors.
 \end{itemize}
% 
% \begin{figure}[h!]
% \centering
% %\includegraphics[width=8.25cm,height=8.25cm]{final_plot_ver3.eps}
% \includegraphics[width=7cm,height=5cm]{Figures/sammonmap6.eps}
% \caption{\small {Two dimensional visualization of speech-music data with {\small {SMCC}} feature using Sammon mapping}}
% \label{samonfig}
% \end{figure}

\begin{table}
\centering
\caption{ Confusion matrix for Multipitch Environment Task. (SP-1 denotes speech with single speaker, SP-2 denotes speech mixture
with two speakers and so on). Class wise accuracy is given as the last column entry.}
\scalebox{1.0}{\begin{tabular}{l|l|c|c|c |c|c|}
\multicolumn{2}{c}{}&\multicolumn{2}{c}{}\\
\cline{3-7}
\multicolumn{2}{c|}{}&SP-1&SP-2& \multicolumn{1}{|c|}{SP-3}& \multicolumn{1}{|c|}{SP-4}& \multicolumn{1}{|c|}{\%}              \\
%\multicolumn{2}{c|}{}&Positive&Negative& \multicolumn{2}{c|}{}&Positive&Negative&\\
\cline{2-7}
& SP-1 & $26$  & $0$ & $0$ & $0$ & $100$ \\
\cline{2-7}
& SP-2 & $2$  & $20$ & $3$ & $1$  & $77$     \\
\cline{2-7}
& SP-3& $0$  & $5$ & $17$ & $4$   & $65$ \\
\cline{2-7}
& SP-4 & $0$  & $3$ & $8$ & $15$  & $58$        \\
\cline{2-7}
\end{tabular}}
\label{multi}
\end{table}
A multi-
dimensional scaling technique, Sammon mapping \citep{sammon} is
used to visualize the separability of SMCC features in Figure \ref{sammon}. Sammon
mapping is a non-linear mapping of high dimensional feature
vectors to low dimensional space based on gradient search.
In the figure, SMCC features computed for a single speaker (SP-1)  and a speech mixture (SP-3) are plotted.
% The iterative EM algorithm is used to estimate the parameters of each
% Gaussian component and the mixture weights \citep{munoz}.
In the proposed method, 20 dimensional SMCC feature vectors are computed in the front-end using the steps described above. A Gaussian Mixture Model (GMM) based classifier is used in 
the classification stage.
The feature vectors computed from the training set are used to build  models for one-speaker case, two speakers case and so on.  
Out of 180 files available in the dataset, 60\% files are used for training and the rest for testing. 12 component Gaussian mixture
models {\small{(GMM)}} are used in the modelling different classes of the speech mixtures.  
During the testing phase, the classifier evaluates the likelihoods of the unknown speech mixture data against these models.   The model that gives  the maximum accumulated likelihood is 
declared as the correct match. The performance of the  aforesaid feature was evaluated on speech mixtures generated by the subset of GRID dataset \citep{grid1}. 
%  {\small{GRID}} \citep{grid1} is a large multi talker audiovisual sentence corpus to support joint computational-behavioral studies in speech perception.  The corpus
% consists of high-quality audio and video  recordings of 1000 sentences spoken by each of 34 talkers (18 male, 16 female).  Audio samples are sampled
% at sampling rate of 25000 Hz.  
%A subset of 40 audio files are used  for generating mixtures for the evaluation.
%Each audio mixture is processed using a Hamming window with a frame length of 30 ms and hop size of 10 ms.
% The test was conducted on speech mixtures, generated with 0 dB Target to Interference Ratio (TIR) which is considered the most 
% difficult situation in estimating the number of speakers.
The results are tabulated as a confusion matrix in  Table \ref{multi}. The overall accuracy is 75 \%. All the single speaker test utterances are classified correctly.  The  results show that the proposed feature
is a promising one in estimating number of speakers in a mixed speech.
%Over

\section{Conclusion}
\label{sec:conclusion}
A  phase based approach  for multipitch estimation is presented in this paper,  yielding
competitive performance as compared to other state  of the art approaches. In the proposed algorithm, the power spectrum is first flattened in order 
to annihilate the system characteristics.
The flattened spectrum is processed using {\small{{MODGD}}}  algorithm to estimate the predominant pitch in each frame in the first pass.
Then the estimated pitch and its harmonics are filtered out using  comb filter. In the second pass, the residual spectrum is again analysed using the group delay algorithm 
to estimate the second candidate pitch. The pitch grouping stage followed by the post processing step results in final pitch trajectories.
The performance of the proposed algorithm was evaluated on speech mixtures with cross gender (Female, Male), same gender (Male/Male, Female/Female) patterns on versatile datasets. The remarkable point in the proposed
method is that  the proposed method is an unsupervised approach using phase information. It does not require pre-training on source models from isolated recordings.
 The problem of estimation of 
number of speakers in a speech mixture is also addressed using a variant of group delay feature, Source-MODGD Cepstral Coefficient features and
evaluated the performance using a subset of GRID corpus. The results obtained in the multipicth experiments show that the proposed algorithm is promising one in multipitch environment for real audio recordings.

\end{sloppypar}

\section {Acknowledgement}
\label{ack}
The authors would like to thank M. Wu, W. Jin, D.L.Wang and Guy J. Brown for sharing their algorithm for multipitch estimation.

\bibliographystyle{elsarticle-harv}

\bibliography{references2_ncc_new.bib}
%\end{sloppy}
\end{document}